\documentclass[12pt]{elsarticle}

\usepackage[american]{babel}
\usepackage{amsmath}
\usepackage{amssymb}
\usepackage{bm}
\usepackage{hyperref}
\usepackage{graphicx}
\usepackage{url}
\usepackage{multirow}
\usepackage{booktabs}
\usepackage[super]{nth}
\usepackage{siunitx}

\journal{Probabilistic Engineering Mechanics}

\begin{document}

\begin{frontmatter}

\title{A framework for probabilistic prediction of remaining useful life in structural materials}

\author[UERJ,CEPEL]{Victor Maudonet}
\author[CEPEL]{Carlos Frederico Trotta Matt}
\author[UERJ,LNCC]{Americo Cunha~Jr\corref{mycorrespondingauthor}}
\cortext[mycorrespondingauthor]{Corresponding author}
\ead{americo@lncc.br}

\address[UERJ]{Universidade do Estado do Rio de Janeiro -- UERJ, Rio de Janeiro, Brazil}
\address[CEPEL]{Centro de Pesquisa de Energia Elétrica -- CEPEL, Rio de Janeiro, Brazil}
\address[LNCC]{Laboratório Nacional de Computação Científica -- LNCC, Petrópolis, Brazil}

\begin{abstract}
Accurate prediction of remaining useful life under creep conditions is essential for the structural reliability of high-temperature components in critical engineering systems. Traditional approaches based on deterministic parametric models often overlook the substantial variability inherent in experimental data, compromising the accuracy and robustness of long-term predictions. This study introduces a probabilistic framework to quantify uncertainties in creep rupture time prediction. Robust regression techniques are first applied to mitigate the influence of outliers and enhance the stability of model estimates. Global sensitivity analysis using Sobol indices is then employed to identify the dominant contributors to model uncertainty, followed by Monte Carlo simulations to propagate these uncertainties and estimate the distribution of the remaining useful life. Finally, model selection is guided by statistical criteria, including the Akaike and Bayesian information criteria, to identify the most reliable predictive model. The proposed framework not only enables the definition of safe operational limits with quantifiable confidence levels but is also general and extensible to other time-dependent degradation phenomena, such as fatigue and creep-fatigue interaction. 
\end{abstract}

\begin{keyword}
structural materials \sep creep modeling \sep remaining life \sep uncertainty quantification \sep Sobol indices \sep Monte Carlo simulation
\end{keyword}

\end{frontmatter}

\section{Introduction}

This research contributes to the growing field of predictive models estimating the remaining useful life (RUL) of components operating under high temperatures, where time-dependent damage mechanisms such as creep and fatigue play critical roles. Among various techniques, time-temperature-parameter (TTP) methods are instrumental in the design and management of high-temperature components. Typically, such methods are implemented through deterministic approaches, as recommended by design codes such as the ASME Boiler and Pressure Vessel Code for metallic components in high-temperature conditions \cite{Dowling2012,Dias2023p185,Evans2011p2838,Gao2021p435,Kamdem2023p624,Wang2020p37}. However, deterministic models and structural integrity codes often overlook aleatory and epistemic uncertainties arising from measurement errors and modeling inaccuracies in predictive creep data and models. These uncertainties are yet to be systematically addressed and still largely depend on subjective judgments \citep{Roya2010,Smith:2013,soize2017}. Despite advancements in mathematical modeling and extrapolation techniques that incorporate microstructural variables, the significant dispersion in creep data, coupled with the complexities of mathematical modeling and a limited understanding of underlying failure mechanisms, often results in substantial discrepancies in creep life predictions \citep{Phanetal2017,Praveenetal2019,Zentuti2017}.

Although creep and fatigue are often treated separately, modern applications in the aerospace, nuclear, and energy sectors frequently encounter service conditions in which both mechanisms coexist or interact. The transition from deterministic to probabilistic life prediction is therefore not only relevant for creep modeling but also directly applicable to fatigue and creep-fatigue interaction \cite{Gu2022p106677,Nie2025p108732}. In both phenomena, damage accumulates over long service times and is sensitive to uncertainties in material properties, loading conditions, and environmental influences. Probabilistic frameworks originally developed for creep, such as the one proposed here, can thus inform generalized RUL estimation strategies in complex environments with multiple degradation mechanisms.

Given these challenges, there is a growing need to shift from deterministic to probabilistic methodologies that formally account for uncertainties for robust design and reliability assessment. Such a transition becomes even more pressing in the current context, where life-extension strategies must reduce conservatism without compromising safety. Reliability-based design and maintenance planning require quantifying uncertainties and evaluating failure probabilities beyond the scope of traditional deterministic models.

Past research in probabilistic methods for predicting creep remaining useful life has been relatively sparse. Notable examples include the work of Lou {\it et al.}\ \cite{MohammadLouetal2020}, who developed a probabilistic Larson-Miller model, and Zhao {\it et al.}\ \cite{zhaoetal2009}, who introduced a service condition-creep rupture property interference (SCRI) model. More recently, fatigue researchers have emphasized the need for probabilistic fatigue life models \cite{Dias2019,AI2019165,Liu2023p107734,Meggiolaro2023p107315,Ball2024p108569} and data-driven machine learning frameworks \cite{Dabetwar2021p021004,Gu2022p106677,Deng2025p108647}. However, these methodologies often omit key steps such as outlier elimination, global sensitivity analysis, or the systematic quantification of input uncertainty.

In this context, we propose a probabilistic framework for quantifying uncertainty and predicting high-temperature creep in metallic components. While the approach is demonstrated using a time-temperature-parameter method for creep rupture, its structure is general and extensible to fatigue and other parametric life prediction models. Our methodology incorporates robust regression to account for data irregularities, Sobol-based global sensitivity analysis to rank influential parameters, and Monte Carlo simulations to propagate uncertainty. It shares conceptual foundations with the probabilistic treatment proposed by Zhang {\it et al.}\ \cite{Zhangetal2024} but introduces enhanced mathematical rigor, parameter correlation modeling, and formal model selection using information-theoretic criteria.

The resulting framework is code-agnostic and modular, enabling integration with more sophisticated constitutive models, including continuum damage mechanics-based formulations \cite{HarlowDelph1995,HossainStewart2021} or fractional-calculus-based \cite{Ribeiro2021p1184,TellesRibeiro2025p20240861}. Ultimately, this research aims to justify a paradigm shift toward probabilistic methodologies in structural integrity analysis—identifying sources of uncertainty, evaluating the performance of statistical techniques, and quantifying their added value over deterministic procedures in predictive lifing of critical systems.

\section{Parametric models for creep prediction}
\label{sec2}

In this section, we outline the foundations of three widely-used parametric models for creep prediction: the Larson-Miller (LM), Orr-Sherby-Dorn (OSD), and Manson-Succop (MS) models. This overview, which presents a summary of these models rather than a comprehensive discussion, is intended for clarity and completeness. Although these models address creep, their parametric structure parallels that of fatigue life modeling (e.g., stress-life or strain-life curves), allowing extension of the framework to fatigue domains.

\subsection{The Larson-Miller Model}

The Larson-Miller (LM) model \citep{MohammadLouetal2020,Miller1952,Choudharyetal2014,Ayubalietal2021} is a common parametric method used to predict the rupture time of metals under creep. Deriving from the Arrhenius relation at a constant stress, the LM model considers a variable temperature $T$ and creep activation energy $Q_c$, resulting in the following mathematical relation
\begin{equation}
P_{LM}(\sigma) = T \, \left(C_{LM} + \log{t_r} \right) \, ,
\label{eq:1}
\end{equation}
where $C_{LM}$ and $P_{LM}$ are the Larson-Miller constant and parameter, respectively. The $P_{LM}$ parameter allows for the superposition of rupture curves into a single master curve, a graphical representation of the logarithm of rupture time ($\log{t_r}$) plotted against $1/T$ at a constant stress $\sigma$ \citep{Dowling2012,Monkman1956}.

While the LM model offers an effective means of predicting creep rupture time --- by isolating $t_r$ in the previous formula
\begin{equation}
t_r = 10^{P_{LM}(\sigma)/T - C_{LM}} \, ,
\end{equation}
it suffers from limitations in terms of physical realism, particularly due to variations in the $C_{LM}$ constant across different alloys and processing conditions \citep{Gilbert2007}. To improve reliability and account for these uncertainties, it is suggested to apply this model within a probabilistic framework as argued by Mohammad Lou {\it et al.} \cite{MohammadLouetal2020}.

\subsection{The Orr-Sherby-Dorn Model}

The Orr-Sherby-Dorn (OSD) model \cite{MohammadLouetal2020,Ayubalietal2021,Orr1954} presents a unique approach where the constant in the LM equation becomes a function of stress, and the LM parameter becomes a constant. This rearrangement of the LM relation leads to the OSD equation
\begin{equation}
P_{OSD}(\sigma) = \log{t_r} - \frac{C_{OSD}}{T} \, ,
\label{OrrSherbyDorn}
\end{equation}
where $P_{OSD}$ and $C_{OSD}$ are the Orr-Sherby-Dorn parameter and constant, respectively. A key assumption of the OSD model is that the creep activation energy, $Q_c$, remains constant over the creep curve, an assumption with limited empirical support \cite{Dowling2012,Orr1954,Carreker1950}. The OSD model's limitations become evident when structural instabilities and multiple rate processes are involved \cite{Dowling2012,Mullendore1963,Allen1960}. As with the LM model, these limitations suggest the need for a probabilistic approach that can accommodate uncertainties and provide more reliable predictions.

\subsection{The Manson-Succop Model}

The Manson-Succop (MS) model \cite{Manson1956}, based on an analysis of the iso-stress lines in a log-time versus temperature plot, defines its parameter $P_{MS}$ through the parallelism of these lines. The MS equation is given by
\begin{equation}
P_{MS}(\sigma) = \log(t_r) + C_{MS} \, T \, ,
\label{eq:3}
\end{equation}
where $P_{MS}$ and $C_{MS}$ are the Manson-Succop parameter and constant, respectively. While the MS model, similar to the LM and OSD models, provides a useful tool for predicting creep life, its deterministic nature often leads to overestimations and errors, particularly for long-term creep life predictions \cite{Dowling2012,Zharkova2003}. Hence, a probabilistic approach is suggested for more accurate and safer creep life predictions.

\subsection{Limitations of parametric models}

The effectiveness of parametric models for creep life prediction, including the LM, OSD, and MS models, is well-documented and acknowledged \cite{Abdallahetal2018}. Their wide application stems from their ability to reduce costs and timelines for collecting long-term creep data. By extrapolating short-term creep data using a time-temperature parameter, these deterministic models provide a procedure for generating a single ``master curve''. This curve, where stress is plotted against an empirical parameter that combines time and temperature, is constructed from available short-term creep data, and predictions for longer durations can be extrapolated from it \cite{Abdallahetal2018}. Although more mathematically simple than their continuum damage mechanics-based counterparts, these parametric methods have a great advantage, at least in theory, of requiring only a
relatively small amount of data to establish the required master curve \cite{Abdallahetal2018}. 

Despite their recognized utility and proved validity for long-term creep predictions, these models are not without limitations. There are several areas where these parametric models may not fully capture the complexities and uncertainties inherent in real-world scenarios, making their outputs less reliable under certain circumstances, as acknowledged by Refs. \cite{Abdallahetal2018,Zhangetal2024}.

The first major limitation lies in the models' deterministic nature. They inherently assume constant material properties and unvarying loading conditions. In reality, however, both operational environments and material properties fluctuate significantly due to numerous factors. For example, manufacturing processes can introduce variations in a material's properties. Additionally, material inhomogeneities in real-world scenarios can contribute to variances that these models cannot account for. Operational uncertainties, such as temperature fluctuations or irregular loading conditions, can also impact the creep behavior of materials, introducing additional sources of discrepancies between model predictions and actual outcomes \cite{Yang2025,Gong2024,Gharaibeh2024}.

Secondly, these models have been observed to struggle when tasked with the prediction of long-term creep life. These models are based on extrapolation of short-term creep test data obtained for larger stress levels. While useful for reducing testing time, they do not take into account that microstructural damage mechanisms induced by larger stress levels may be quite different from those verified in an actual component subjected to much lower stress levels. Hence, larger discrepancies are commonly verified when extrapolating creep life using these deterministic models, as both model-parameter uncertainties and model uncertainties induced by different microstructural damages evolving at unknown time rates are not taken into account. This dispersion means that different predictions frequently arise, even when using experimental data acquired under apparently identical macroscopic conditions. The introduction of these uncertainties in long-term forecasts can cause the models to become less accurate. The inaccuracy is often compounded in high-stress, high-temperature applications, where the creep behavior becomes notably nonlinear.

Moreover, it is worth noting that these parametric models often ignore changes in creep fracture mechanisms across different temperature ranges and fracture durations. This can lead to inaccuracies and overestimations in the projected long-term creep life. This is particularly evident in the MS model which assumes fixed values of constants over a wide range of temperatures and fracture durations.

Lastly, these models tend to oversimplify the processes they are trying to predict. Complex interactions between multiple factors affecting the creep strength of alloys, particularly at high temperatures, are often not adequately accounted for. This simplification becomes especially problematic in scenarios involving structural instabilities or where multiple rate processes are at play, such as in complex alloys.

Considering these limitations, it is necessary to adopt a probabilistic approach in order to account for the inherent uncertainties in creep life prediction. The probabilistic approach not only provides a risk-based perspective, but also optimizes design, maintenance, and operation strategies. Adopting such an approach is crucial during the initial design phase of high-temperature equipment to ensure a specific service life as determined by code. This strengthens the argument for the application of parametric models from a probabilistic perspective, leading to a more reliable and accurate prediction of creep life.

%
\section{Uncertainty quantification framework}

\begin{figure}
\centering
\includegraphics[width=0.95\textwidth]{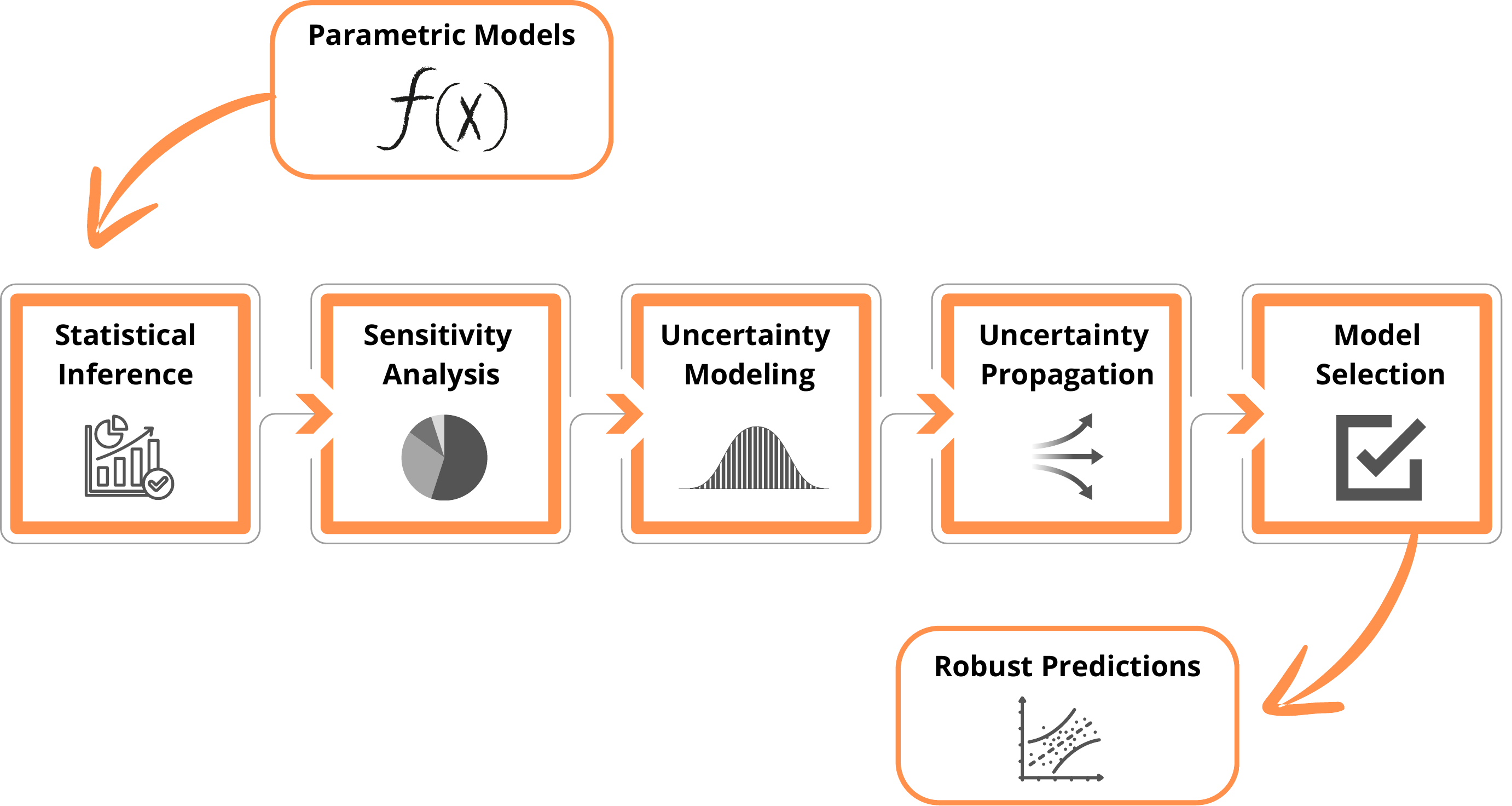}
\vspace{2mm}
\caption{Schematic representation of the proposed probabilistic uncertainty quantification framework for predicting remaining useful life (RUL). The framework integrates robust statistical inference, global sensitivity analysis, uncertainty modeling and propagation, and model selection. Designed for time-dependent degradation processes such as creep and fatigue, the methodology enables quantification of lifetime variability and supports risk-informed decision-making for structural reliability and predictive maintenance.}
\label{fig:Uq_framework}
\end{figure}

The uncertainty quantification (UQ) framework proposed here, inspired by the works \citep{Dias2019}, \citep{Dias2018}, and \citep{Nispel2021}, encompasses a five-step process. This process, schematically illustrated in Figure \ref{fig:Uq_framework}, includes stages of statistical inference, sensitivity analysis, uncertainty modeling, uncertainty propagation, and model selection. Each of these steps is explored in greater detail below. Although this framework is demonstrated for creep models, the same sequence can be applied to fatigue formulations, including stress-based and strain-based life estimation models, or more generally, to any kind of parametric prediction model $y = f(\bm{x})$, which represents the relationship between a quantify of interest $y$ and a set of parameters encapsulated into a vector $\bm{x}$. For instance, for the models of section~\ref{sec2} we have $y = t_r$ and $\bm{x} = (P_{\square}, C_{\square})$, where $\square \in \{LM, OSD, MS\}$.

\subsection{Statistical inference}

The first phase in our uncertainty quantification framework involves defining the mathematical relationship between empirical parameters of deterministic creep models ($P_{LM}$, $P_{OSD}$, $P_{MS}$) and the mechanical stress, $\sigma$. The initial process involves data pre-processing to minimize the influence of potential outliers that could potentially distort the analysis. To this end, we apply the winsorization method, which has proven to be effective in controlling extreme values \cite{Boyd2018}. This technique operates by assigning all outliers to a specific percentile of the data. For instance, a 95\% winsorization would set all data points below the 5th percentile to the value at the 5th percentile, and all data above the 95th percentile to the value at the 95th percentile.

Winsorization is adopted here as a preprocessing step due to its simplicity and its ability to limit the influence of extreme values without modifying the sample size or introducing weighting schemes. This approach remains decoupled from the regression procedure, which is advantageous in the present multi-stage framework. Sensitivity checks with moderate variations of the winsorization threshold showed negligible impact on the selected polynomial structure and on subsequent probabilistic predictions.

After the data has been winsorized, we proceed with identifying the relationship between the creep model parameter $P_{\square}$ and the stress $\sigma$, i.e., 
\begin{equation}
P_{\square} = a_0 + a_1 \, \sigma + \cdots + a_{n-1} \, \sigma^{n-1} \, .
\label{eq:sigma-poly}
\end{equation}
We approach this problem from a data-driven perspective, employing a specialized form of sparse regression, the sequential threshold least-square (STLS), along with cross-validation \citep{Brunton2016,brunton2022}. The use of STLS allows for the identification of polynomial relationships between variables, free from assumptions regarding the polynomial degree. In mathematical terms, given a design matrix $[\bm{A}] \in \mathbb{R}^{m \times n}$,  the observations $\bm{y} \in \mathbb{R}^{m}$, and the unknown vector $\bm{x} \in \mathbb{R}^{n}$, 
\begin{equation}
[\bm{A}] = \begin{bmatrix}
1 & \sigma_1 & \sigma_1^2 & \cdots & \sigma_1^{n-1} \\
1 & \sigma_2 & \sigma_2^2 & \cdots & \sigma_2^{n-1} \\
\vdots & \vdots & \vdots & \ddots & \vdots \\
1 & \sigma_m & \sigma_m^2 & \cdots & \sigma_m^{n-1}
\end{bmatrix} \, ,
~~
\bm{y} = 
\begin{bmatrix}
P_{\square ~ 1}\\
P_{\square ~ 2}\\
\vdots\\
P_{\square ~ m}
\end{bmatrix} \, ,
~~
\bm{x} = 
\begin{bmatrix}
a_{0}\\
\vdots\\
a_{n-1}
\end{bmatrix} \, ,
\end{equation}
the STLS solves the optimization problem
\begin{equation}
\bm{x}^{\ast} = \arg\min_{ \bm{x} ~ \in ~ \mathbb{R}^n}  ~ || \, \bm{y} - [\bm{A}] \, \bm{x} \, ||_2^2 + \lambda \, || \, \bm{x} \, ||_0  \, ,
\label{eq:Regression}
\end{equation}
where $||.||_2$ denotes the $l_2$-norm, $||.||_0$ is the ``$l_0$-norm'' denoting the number of nonzero entries in a vector, and $\lambda$ is a sparsity-inducing regularization parameter. A parsimonious polynomial model emerges as a result of enforcing sparsity on the regression coefficients, eliminating redundant powers, and determining the optimal fit function based on a rational criterion.

It should be noted that the sparsity constraint in the STLS procedure may filter out weak nonlinear terms, particularly in noisy datasets. In this work, such behavior is intentional, as the objective is to obtain parsimonious models that prioritize predictive robustness. To ensure that retained terms are consistently supported by the data, we use a cross-validation procedure, thereby reducing the risk of overfitting and improving the stability of subsequent uncertainty quantification.

Cross-validation is utilized to evaluate each polynomial's predictive ability and to select the optimal model using the root mean squared error (RMSE) criterion. The cross-validation process splits the experimental data into complementary subsets, performs the analysis on one subset (the training set), and validates the analysis on the other subset (the validation or testing set). This procedure is formally described as follows: given a dataset of size $m$, it is divided into $k$ equally sized folds. For each fold $i = 1,...,k$, a model is trained on all data excluding fold $i$ and is then tested on fold $i$, yielding a prediction error $e_i$. The RMSE is then computed as

\begin{equation}
RMSE = \sqrt{\frac{1}{k} \sum_{i=1}^{k} e_i^2} .
\label{eq:RMSE}
\end{equation}

The data is then shuffled, the partitioning process is repeated, and the polynomial yielding the lowest RMSE value is chosen as the optimal model.

\subsection{Sensitivity analysis}

The second stage of our uncertainty quantification framework conducts a rigorous global sensitivity analysis, aiming to discern the primary sources of output variability introduced by the input parameters into parametric predictive models of the form $y = f(\bm{x})$. In the presence of a substantial number of potential input parameters, sensitivity analysis acts as a crucial tool in pinpointing which parameters require a more careful consideration and which ones can be reasonably neglected in the subsequent process \citep{Zentuti2017}.

Unlike classical sensitivity analysis (based on partial derivatives of the response with respect to the inputs), which provides only local information, in global sensitivity analysis it is possible to map the dependence of the model's response with respect to variations of the parameters in value ranges. This is a probabilistic approach, where a randomized version of the parametric model, denoted by $Y = f(\bm{X})$, is considered. Here $\bm{X}$ is a random vector that stores random variables corresponding to stochastic versions of the parametric model input parameters, while Y is a random variable that represents the possible variations in the quantity of interest (QoI), a.k.a. model output.

In this global sensitivity analysis setting, a variance decomposition technique is used to address the dependence of the response with respect to the model's inputs, individually or jointly. This is accomplished by calculating sensitivity indices, quantities that capture the contribution of each input parameter's variability towards the overall output variability. The choice of sensitivity indices for this work lies in Sobol indices, well-regarded for their effectiveness in such tasks \citep{Sudret2008}.

Sobol' sensitivity indices can be expressed as
\begin{equation}
S_i = \frac{\mathbb{V}\left(\mathbb{E}\left(Y \mid X_i\right)\right)}{\mathbb{V}(Y)} \, ,
\label{eq:Sobol}
\end{equation}
where $\mathbb{V}(\cdot)$ and $\mathbb{E}(\cdot)$ are the variance and expectation operations, respectively, $Y$ is the model output, and $X_i$ is the $i$-th input parameter.

Several strategies exist for evaluating sensitivity indices. In this work, we employ both the Monte Carlo (MC) method and Polynomial Chaos Expansion (PCE) \citep{kroese2011,cunhajr2014p1355,ghanem2003,xiu2010}. The Monte Carlo method serves as a reference for the process, while the use of PCE aids in circumventing cancellation errors that might occur during the computation of higher-order indices, leveraging the method's relatively low computational cost \citep{Sudret2008,UQLab}.

For the MC method, we generate 10,000 samples, ensuring statistical convergence for the rupture time prediction. On the other hand, for the PCE approach, the expansion coefficients are determined using the ordinary least squares method over 1,000 samples and with a maximum degree of 10 for the polynomial expansion.

The model's output $Y$ is expanded over a set of multivariate orthogonal polynomials $\Psi_i(\mathbf{X})$ to compute PCE
\begin{equation}
Y(\mathbf{X}) = \sum_{i=0}^{+\infty} \alpha_i \, \Psi_i(\mathbf{X}) \, ,
\label{eq:PCE}
\end{equation}
where $\alpha_i$ are the unknown coefficients to be determined. These coefficients are determined by projecting $Y$ onto each basis function, i.e.,
\begin{equation}
\alpha_i = \frac{\mathbb{E}( Y \, \Psi_i )}{\mathbb{E}( \Psi_i^2)} \, ,
\label{eq:PCEcoeff}
\end{equation}
where the expectation operator plays the role of an inner product.

The total variance $V$ and the conditional variances $V_i$ can be expressed in terms of the PCE coefficients 
\begin{equation}
V = \sum_{i=1}^{+\infty} \alpha_i^2, \quad V_i = \sum_{j \in \mathcal{I}_i} \alpha_j^2 \, ,
\label{eq:variances}
\end{equation}
where $\mathcal{I}_i$ is the set of multi-indices corresponding to all polynomials that only depend on $X_i$. The first-order and total-order Sobol indices are then computed by the ratios of partial variances to the total variance
\begin{equation}
S_i = \frac{V_i}{V}, \quad S_{T_i} = 1 - \frac{V_{-i}}{V} \, ,
\end{equation}
where 
\begin{equation}
V_{-i} = \sum_{j \notin \mathcal{I}_i} \alpha_j^2 \, .
\end{equation}

Lastly, we assume the model input parameters to be independent, uniformly distributed random variables with physically justifiable lower and upper bounds. The uniform distribution for each parameter $X_i \sim \mathcal{U}(a_i, b_i)$, where $a_i$ and $b_i$ are the lower and upper bounds for the i-th parameter. This assumption simplifies the computational process while providing robustness in the sensitivity analysis.

\subsection{Uncertainty modeling}

The third stage of the uncertainty quantification (UQ) framework involves modeling uncertainties, which are fundamental to understanding the statistical variability of input parameters. This involves determining the probability density function (PDF) for each input parameter, which must be chosen using rational criteria to prevent assumptions from violating the underlying physics of the creep phenomena \citep{soize2017,cunhajr_ekwaro2016}.

Note that the stochastic nature of the input parameters depends on their inherent characteristics and the manner in which creep tests are conducted. In the present study, temperature and mechanical stress are treated as deterministic variables, as they are controlled quantities in standard creep experiments, maintained at prescribed levels with comparatively low variability relative to material response. This modeling choice is deliberate: it isolates the dominant sources of uncertainty associated with material behavior and model parameters, which are the primary focus of the proposed framework.

From a methodological standpoint, extending the framework to account for variability in temperature and stress is straightforward, as these quantities can be incorporated directly as additional random variables during the uncertainty propagation stage. Such an extension would primarily lead to greater dispersion in the predicted rupture time, without altering the framework's structure.

On the contrary, the polynomial coefficients, which define the relationship between the empirical parameters ($P_{LM}, P_{OSD}, P_{MS}$) and the mechanical stress $\sigma$, as well as the empirical constant for each model ($C_{LM}$, $C_{OSD}$, and $C_{MS}$), are treated as random variables. By invoking the Central Limit Theorem, we can presume that these estimators follow an asymptotic Gaussian distribution. Consequently, a multivariate Gaussian distribution is selected to describe their joint statistics.

Consider a random vector $\bm{X}$, where each component represents a polynomial coefficient or an empirical constant. The joint PDF for these components, following a multivariate Gaussian distribution, can be represented as
\begin{equation}
p_{\bm{X}}(\bm{x};\bm{\mu}, [\bm{\Sigma}]) = \frac{1}{(2\pi)^{\frac{d}{2}} \, \sqrt{\det{[\bm{\Sigma}]}}} \exp\left\{-\frac{1}{2}(\bm{x} - \bm{\mu})^T \, [\bm{\Sigma}]^{-1} \, (\bm{x} - \bm{\mu})\right\} \, ,
\label{eq:2}
\end{equation}
where $d$ is the dimension of the random vector $\bm{X}$ (obtained after global sensitivity analysis); $\bm{\mu}$ and $[\bm{\Sigma}]$ denote the mean and covariance matrix of the random vector ${\bf X}$, respectively. The mean vector $\bm{\mu}$ is the least-squares estimator of the input parameters, and the covariance matrix $[\bm{\Sigma}]$ is derived from
\begin{equation}
[\bm{\Sigma}] = \sigma^2_{e} \, ([{\bf A]}^{T} [{\bf A}])^{-1} \, ,
\end{equation}
where $\sigma^2_{e}$ is the variance of the prediction error vector, defined as the discrepancy between estimated and the observed creep rupture times for available observations data \citep{Smith:2013}. Here, $[{\bf A}]$ refers to the well-known design matrix appearing in the least-squares mathematical formulation \citep{Boyd2018}.

The predictor map relies on both the independent variable $\sigma$ and the model parameters $\bm{x}$, i.e., $y(\sigma^{(k)},\bm{x})$, and the component $A_{kj}$ of the design matrix is given by the following partial derivative
\begin{equation}
A_{kj} = \frac{\partial{y}}{\partial x_{j}}(\sigma^{(k)},\bm{x}) \, ,
\end{equation}
evaluated at a given set of model parameters, say at $\bm{x} = \bm{\mu}$.

The model-parameter uncertainty is fully characterized by the multivariate Gaussian PDF given by Eq. (\ref{eq:2}), the mean vector $\bm{\mu}$ and covariance matrix $[\bm{\Sigma}]$. The final stage in the UQ framework involves quantifying the impact of model-parameter uncertainty on the uncertainty of the QoI computed from each parametric creep model. In this context, the creep rupture time serves as the output QoI.

\subsection{Uncertainty propagation}

The stage of uncertainty propagation entails the generation of independent samples of the input parameter vector ${\bf X}$ from the multivariate Gaussian probability density function specified earlier. These samples are then used to calculate instances of the creep rupture time $t_r$. To execute this stage, we employ the Monte Carlo method, a widely used statistical sampling technique for uncertainty propagation \citep{cunhajr2014p1355,kroese2011}. For this work, we generate 10,000 samples, a number determined through a convergence analysis of the mean value and variance of the creep rupture time.

The Monte Carlo uncertainty propagation process can be compartmentalized into three general steps: pre-processing, processing, and post-processing.

In the pre-processing step, we generate model input samples using the Cholesky decomposition of the correlation matrix, effectively transforming the covariance matrix into a lower-triangular matrix. This allows us to generate a sample vector with the covariance properties of the system being modeled by applying this matrix to a sample vector of uncorrelated Gaussian samples and adding the mean vector of the input parameters.

The processing step involves solving each creep model using the input samples generated in the pre-processing step. This results in the generation of 10,000 samples of the rupture time, $t_r$.

Finally, the post-processing step involves performing several analyses to ensure the derived probability distributions accurately represent the physics of the investigated phenomenon. Here, we generate histograms of the rupture time for each of the four investigated operating conditions for each creep model. We also compute statistical metrics such as the mean, standard deviation, coefficient of variation, skewness, and kurtosis. These metrics serve as descriptive statistics, providing insight into the shape and spread of the distributions generated from the Monte Carlo simulations.

The number of Monte Carlo samples was chosen to ensure stable estimation of the first two statistical moments, which are the primary quantities of interest in reliability assessment. While higher-order statistics such as skewness and kurtosis converge more slowly—especially for heavy-tailed distributions—their role in this work is mainly descriptive. Increasing the sample size would primarily improve the resolution of the distribution tails without significantly affecting the main conclusions.

\subsection{Model selection}

The final phase in the uncertainty quantification framework is model selection, where we endeavor to choose the best probabilistic model that accurately reproduces the experimental data obtained for the four experimental conditions. Our strategy for model selection hinges on statistical measures capable of quantifying both model performance on the training dataset (represented as model fitting error) and model complexity (expressed as the number of uncertain parameters).

To select the optimal model, we employ the Akaike information criterion (AIC) and the Bayesian information criterion (BIC) \citep{wasserman2004,Hastie2016,brunton2022}. AIC is computed as follows
\begin{equation}
{\rm AIC}(\bm{x}) = 2 \, n - 2 \, \ln{L(\bm{x})} \, ,
\label{eq:23}
\end{equation}
where $L(\bm{x})$ signifies the likelihood function, defined as the conditional probability of the observed data (denoted by the vector ${\bf y} \in \mathbb{R}^{m}$) given $\bm{X} = \bm{x}$, with $\bm{x}$ being a specific value of the model parameter vector. That is,
\begin{equation}
L(\bm{x}) \equiv p_{\bm{Y} | \bm{X}}(\bm{y} ~|~ \bm{x}) \, .
\end{equation}

Given that ${\bf y}$ is measured data and therefore fixed, both the likelihood and AIC depend on $\bm{x}$, hence the notation $L(\bm{x})$ and ${\rm AIC}(\bm{x})$.

Assuming an observation model with additive error $\bm{\epsilon} \in \mathbb{R}^{m}$, such that $\bm{y} = \hat{\bm{y}}(\bm{x}) + \bm{\epsilon}$, and also assuming that the error follows a Gaussian distribution with zero mean and constant variance $\sigma^2_e$, the likelihood becomes
\begin{equation}
L(\bm{x}) = \frac{1}{(2 \pi)^{m/2} \, \sigma_e} \, {\rm exp} \left\{-\frac{1}{2\,\sigma^2_e} \left(\bm{y} - \hat{\bm{y}}(\bm{x})\right)^{T} \, \left(\bm{y} - \hat{\bm{y}}(\bm{x})\right) \right\} \, ,
\label{likelihood}
\end{equation}
where $m$ denotes the dimension of the observed data vector $\bm{y}$ and $\hat{\bm{y}}(\bm{x}) \in \mathbb{R}^{m}$ represents the predicted creep rupture time for a given set of model parameters, $\bm{x}$.

Conversely, BIC is defined as
\begin{equation}
{\rm BIC}(\bm{x}) = n \, \ln{m} - 2 \, \ln{L(\bm{x})}.
\label{eq:21}
\end{equation}

Model selection is carried out by calculating AIC and BIC for $\bm{x} = \bm{\mu}$, providing a representative assessment of model fidelity and complexity under the assumed likelihood formulation. The probabilistic parametric creep model with the lowest AIC and BIC is chosen as the best model \citep{Hastie2016}. Note that AIC provides less penalization for complex models compared to BIC, thereby giving more emphasis to model performance and tending to select more complex models \citep{Murphy2012}.

\subsection{Extension to time-dependent degradation models}

Although the proposed framework is formulated for parametric models of the form $y = f(\bm{x})$, it can be naturally extended to time-dependent degradation phenomena. In such cases, the system response is governed by a dynamical model describing the evolution of internal state variables, for example
\begin{equation}
\dot{z}(t) = g\big(z(t),\,u(t),\,\bm{x}\big), \qquad y = h\big(z(t_f)\big),
\end{equation}
where $z(t)$ represents state variables associated with damage evolution, $u(t)$ denotes the loading history, $\bm{x}$ is a vector of parameters, and $y$ is the quantity of interest evaluated at a final time $t_f$.

Within this setting, the UQ framework remains unchanged in structure. The stages of statistical inference, sensitivity analysis, uncertainty modeling, and uncertainty propagation are applied to the dynamical system, with uncertainty affecting both model parameters and system trajectories. In particular, Monte Carlo simulation is performed over realizations of $\bm{X}$ (and, if applicable, $U(t)$), yielding a distribution of responses $Y$.

This perspective highlights the modular nature of the proposed framework, which can be directly coupled with state-dependent or history-dependent models, such as those used for fatigue and creep-fatigue interaction, without modifying its core components.

\section{Results and discussion}

\subsection{Polynomial representation of creep model parameters}

Data for 1CrMoV steel, drawn from the National Institute for Material Science (NIMS) database, underwent a winsorizing procedure to establish a reliable link between the empirical parameters $(P_{LM}, P_{OSD}, P_{MS})$ and the mechanical stress $(\sigma)$ for each respective creep model. The winsorizing method sets data below the \nth{5} percentile to the \nth{5} percentile, and data above the \nth{95} percentile to the \nth{95} percentile.

The Sequential Threshold Least Square (STLS) regression, paired with cross-validation, was applied to the treated data. This approach omits polynomial powers (up to the 8th degree) having an absolute value lower than a set threshold, $\lambda$. Four different $\lambda$ values were tested: two for the Larson-Miller parametric model ($\lambda = 0.1$, $\lambda = 0.01$) and two for both Orr-Sherby-Dorn and Manson-Succop parametric creep models ($\lambda = \num{5e-5}$ and $\lambda = \num{5e-6}$).

To evaluate the predictive power of each polynomial derived from the STLS method, we utilized cross-validation. Data was shuffled and divided into a training set (80\%) and a validation set (20\%). For every cross-validation iteration, the STLS method utilized the training set to estimate coefficients of fitting polynomials up to the eighth degree. The validation set then determined the quality of each polynomial's prediction based on the Root Mean Square Error (RMSE). After 100 cross-validation iterations, 800 polynomials per each creep model were generated and ranked. The polynomial with the lowest RMSE was selected as the best representative for the relationship between the empirical parameter and the mechanical stress. Tables \ref{table:value_coefs_LM}, \ref{table:value_coefs_OSD}, and \ref{table:value_coefs_MS} exhibit the mean value of each polynomial coefficient per regression degree over the 100 cross-validation iterations for the Larson-Miller, Orr-Sherby-Dorn, and Manson-Succop parametric models, respectively.

\begin{table}[h!]
\centering
\footnotesize
\caption{Mean values of polynomial coefficients at each regression degree, derived from all cross-validation iterations for the Larson-Miller model.}
\vspace{2mm}
\begin{tabular}{ccc}
\toprule
Max.    & \multicolumn{2}{c}{Threshold Value}\\ 
 Degree & $\lambda = 0.1$          & $\lambda = 0.01$\\ 
\midrule
1  & $22202.0 - ~12.0\sigma$ & $22205.0 - ~12.0\sigma$\\ 
\midrule
2  & $22190.0 - ~11.9\sigma$ & $21580.0 - ~11.7\sigma$\\ 
\midrule
3  & $21863.0 - ~~4.8\sigma$ & $21876.0 - ~~4.9\sigma - 0.04\sigma^2$\\ 
\midrule
4  & $22502.0 - ~23.9\sigma + 0.13\sigma^2$  & $22484.0 - ~23.2\sigma + 0.13\sigma^2$\\ 
\midrule
5  & $22112.0 - ~~9.3\sigma$ & $22127.0 + ~35.1\sigma - 0.80\sigma^2$\\ 
\midrule
6  & $21160.3 + ~34.5\sigma - 0.78\sigma^2$  & $21158.0 - ~~9.8\sigma - 0.04\sigma^2$\\ 
\midrule
7  & $21152.0 + ~35.0\sigma - 0.79\sigma^2$  & $20873.0 + ~50.9\sigma - 1.12\sigma^2$\\ 
\midrule
8  & $19117.0 + 167.9\sigma - 4.17\sigma^2$  & $18747.0 + 190.3\sigma - 4.70\sigma^2 + 0.06\sigma^3$ \\ 
\bottomrule
\end{tabular}
\label{table:value_coefs_LM}
\end{table}

\begin{table}[h!]
\centering
\footnotesize	
\caption{Mean values of polynomial coefficients at each regression degree, derived from all cross-validation iterations for the Orr-Sherby-Dorn model.}
\vspace{2mm}
\begin{tabular}{ccc}
\toprule
Max. &
\multicolumn{2}{c}{Threshold Value}                                                                         \\ 
 Degree & $\lambda = \num{5e-5}$           & $\lambda = \num{5e-6}$\\ 
\midrule
1  & $-27.5 - 0.020\sigma$                   & $-27.3 - 0.020\sigma$\\
\midrule
2  & $-28.7 - 0.002 \sigma$                  & $-26.3 - 0.002\sigma - 0.00003\sigma^2$\\
\midrule
3 & $-29.7 - 0.020\sigma - 0.0002 \sigma^2$  & $-29.9 - 0.020\sigma - 0.00018\sigma^2$\\
\midrule
4  & $-28.6 - 0.010\sigma - 0.0001\sigma^2$  & $-29.9 - 0.002\sigma - 0.00054\sigma^2$\\
\midrule
5  & $-30.2 - 0.050\sigma - 0.0006\sigma^2$  & $-29.6 - 0.020\sigma - 0.00002\sigma^2$\\ 
\midrule
6  & $-31.0 - 0.080\sigma - 0.0012\sigma^2$  & $-28.7 - 0.020\sigma - 0.00043\sigma^2$\\
\midrule
7  & $-28.3 - 0.070\sigma - 0.0019\sigma^2$  & $-26.6 - 0.130\sigma + 0.00280\sigma^2 - 0.00003\sigma^3$\\ 
\midrule
8  & $-29.4 - 0.005\sigma - 0.0001\sigma^2$  & $-26.0 - 0.170\sigma + 0.00380\sigma^2 - 0.00004\sigma^3$\\ 
\bottomrule
\end{tabular}
\label{table:value_coefs_OSD}
\end{table}

\begin{table}[h!]
\centering
\footnotesize
\caption{Mean values of polynomial coefficients at each regression degree, derived from all cross-validation iterations for the Manson-Succop model.}
\vspace{2mm}
\begin{tabular}{ccc}
\toprule
Max. &
\multicolumn{2}{c}{Threshold Value}                                                                         \\ 
 Degree & $\lambda = \num{5e-5}$          & $\lambda = \num{5e-6}$\\ 
\midrule
1  & $27.8 - 0.020\sigma$                   & $27.9 - 0.020\sigma$\\ 
\midrule
2  & $27.5 - 0.016\sigma - 0.00002\sigma^2$ & $24.8 - 0.011\sigma - 0.00002\sigma^2$\\ 
\midrule
3  & $26.6 + 0.001\sigma - 0.00010\sigma^2$ & $26.6 + 0.001\sigma - 0.00011\sigma^2$\\ 
\midrule
4  & $27.9 - 0.040\sigma + 0.00027\sigma^2$ & $27.8 - 0.030\sigma + 0.00018\sigma^2$\\
\midrule
5  & $27.2 - 0.010\sigma - 0.00007\sigma^2$ & $26.4 + 0.018\sigma - 0.00044\sigma^2$\\
\midrule
6  & $26.6 - 0.010\sigma - 0.00052\sigma^2$ & $25.5 - 0.060\sigma - 0.00110\sigma^2 + 0.000008\sigma^3$\\ 
\midrule
7  & $27.6 - 0.040\sigma - 0.00065\sigma^2$ & $28.9 - 0.100\sigma + 0.00280\sigma^2 - 0.000033\sigma^3$\\ 
\midrule
8  & $23.5 + 0.200\sigma - 0.00620\sigma^2 + 0.00008\sigma^3$  & $24.5 + 0.200\sigma - 0.00460\sigma^2 + 0.000065\sigma^3$\\ 
\bottomrule
\end{tabular}
\label{table:value_coefs_MS}
\end{table}

Two key observations stand out from this process. Firstly, polynomials up to the third degree are selected by applying threshold values $\lambda = 0.01$ and $\lambda = \num{5e-6}$. This result aligns with the behavior noticed in the experimental data when plotting creep parameters against mechanical stress. Secondly, the lowest RMSE values for the Larson-Miller, Orr-Sherby-Dorn, and Manson-Succop parametric models are 45.9, 15.7, and 15.9 respectively. The resulting best-fit polynomials (equations \ref{eq:25}, \ref{eq:26}, \ref{eq:27}) from the selection of 800 (for each parametric creep model) are:
\begin{equation}
P_{LM}(\sigma) = 22205 - 12\sigma
\label{eq:25}
\end{equation}
\begin{equation}
P_{OSD}(\sigma) = -26.3 - 0.002\sigma - 0.00003\sigma^2
\label{eq:26}
\end{equation}
\begin{equation}
P_{MS}(\sigma) = 24.8 - 0.011\sigma - 0.00002\sigma^2
\label{eq:27}
\end{equation}

These relationships closely resemble affine functions. Any quadratic terms have coefficients that are very small, implying a limited influence on the overall models.

\subsection{Key parameters of each creep parametric model}

\begin{figure}
\centering
\includegraphics[scale=0.45]{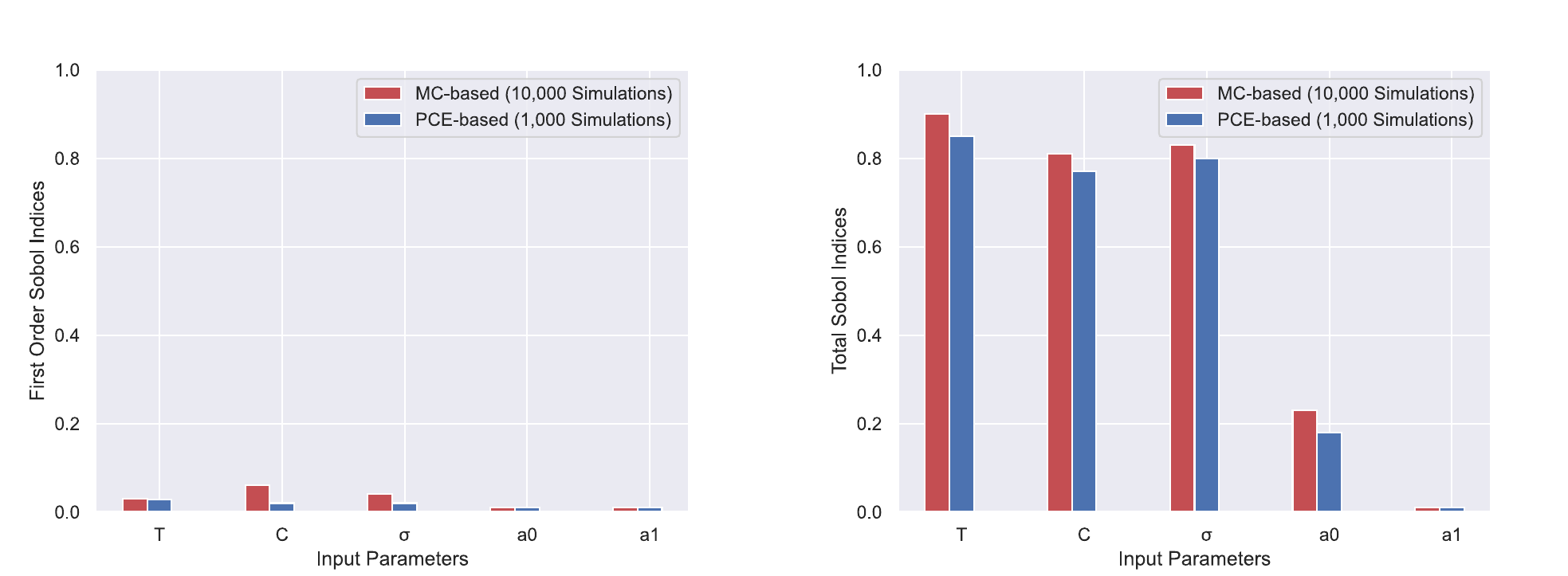}
\vspace{2mm}
\caption{Larson-Miller model's first-order (left) and total (right) Sobol indices.}
\label{fig:sobol_indices_LM}
\end{figure}

\begin{figure}
\centering
\includegraphics[scale=0.45]{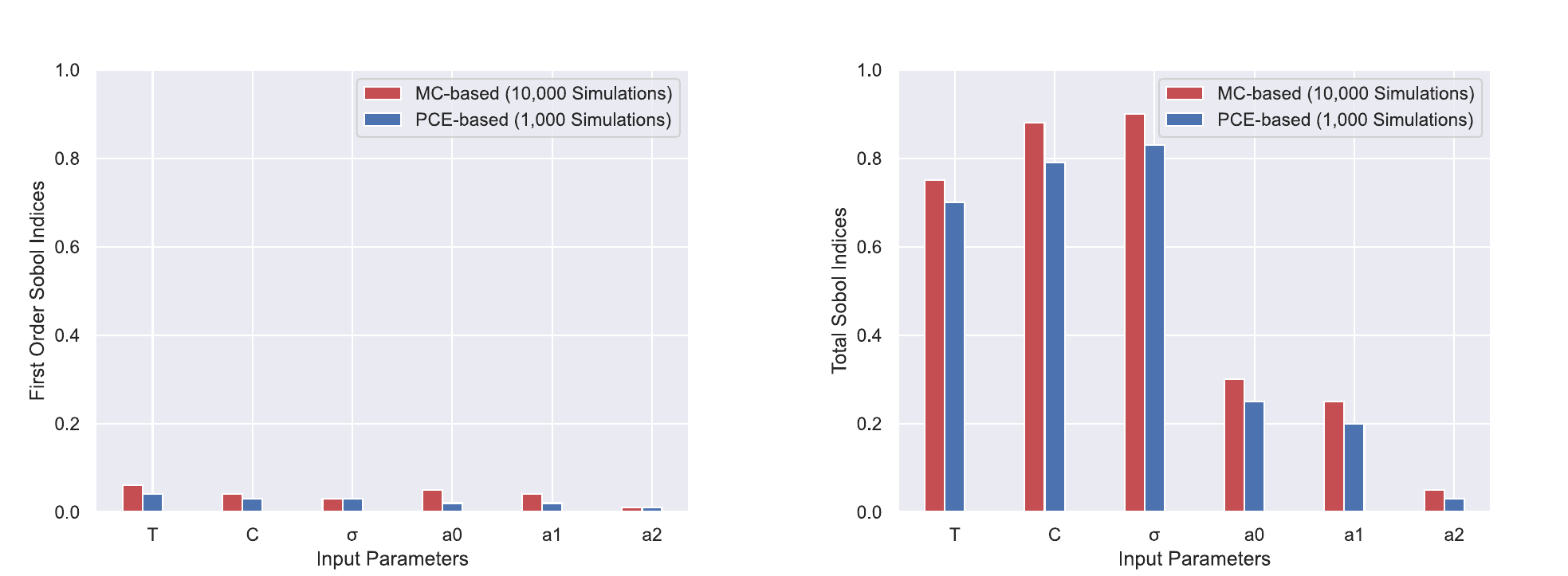}
\vspace{2mm}
\caption{First-order (left) and total (right) Sobol indices for the Orr-Sherby-Dorn model.}
\label{fig:sobol_indices_OSD}
\end{figure}

\begin{figure}
\centering
\includegraphics[scale=0.45]{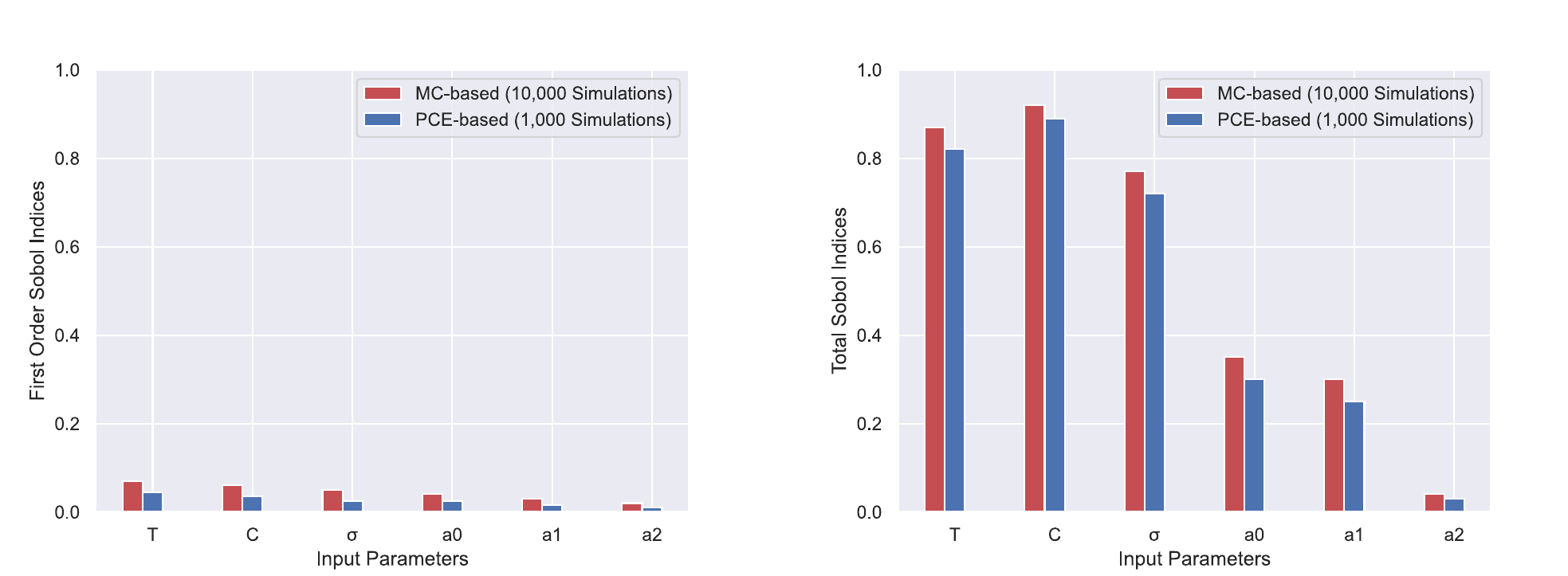}
\vspace{2mm}
\caption{Manson-Succop model's first-order (left) and total (right) Sobol indices.}
\label{fig:sobol_indices_MS}
\end{figure}

The global sensitivity analysis results for each creep parametric model are presented in Figs.\ \ref{fig:sobol_indices_LM}, \ref{fig:sobol_indices_OSD}, and \ref{fig:sobol_indices_MS}. These figures depict the first-order and total Sobol's indices derived from both MC and PCE methods.

Several insights are noteworthy. First, the MC and PCE results are in strong alignment, validating the reliability of the Sobol's indices. Notably, the creep rupture time is predominantly influenced by parameters $a_0$, $\sigma$, $T$, and $C$ across the Larson-Miller, Orr-Sherby-Dorn, and Manson-Succop models, with $a_1$ emerging as another significant factor.

Contrastingly, incremental shifts in the input parameters do not substantially alter the creep rupture time, as reflected by the minor values of the first-order Sobol's indices, all below 0.1. Nevertheless, the interplay of these parameters can drastically sway the creep rupture time.

Total Sobol's indices for parameters $a_1$ and $a_2$, in the Larson-Miller model (Fig.\ \ref{fig:sobol_indices_LM}), and in both Orr-Sherby-Dorn and Manson-Succop models (Figs.\ \ref{fig:sobol_indices_OSD} and \ref{fig:sobol_indices_MS}), are relatively insignificant. This infers a marginal contribution of these parameters to the variance of creep rupture time. Thus, these parameters can be approximated as deterministic values, effectively reducing the complexity of the subsequent probabilistic models.

\subsection{Creep parameters' probabilistic modeling}

Following the global sensitivity analysis, the Central Limit Theorem was utilized to develop a probabilistic model for each parametric creep model's most impactful parameters. These parameters were consequently modeled as jointly Gaussian. Their comprehensive probabilistic description is hence provided by the mean vector $\bm{\mu}$ and the covariance matrix $[\bm{\Sigma}]$.

The estimates for $\bm{\mu}$ and $[\bm{\Sigma}]$ were derived using the STLS method post the cross-validation process. For the Larson-Miller model, denoted as $\bm{X} = [a_0, a_1, C]^{T}$, the values were calculated as follows
\begin{eqnarray}
\bm{\mu}^{\rm LM} &=& [26000, , -9.3, , 23]^{T} \, ,\\
\bm{\Sigma}^{\rm LM} &=&
\begin{bmatrix}
0.005 & 1.700 & 4.100 \\
1.700 & 0.070 & 1.400 \\
4.100 & 1.400 & 3.300
\end{bmatrix} \times \num{e-3} \, .
\end{eqnarray}

For the Orr-Sherby-Dorn model ($\bm{X} = [a_0, a_1, a_2, C]^{T}$), the results were
\begin{eqnarray}
\bm{\mu}^{\rm OSD} &=& [-26.3, , -0.0016, , -0.000034, , 21000]^{T} \, ,\\
\bm{\Sigma}^{\rm OSD} &=& \begin{bmatrix}
3.100 & 2.500 & 5.300 & 3.600\\
2.500 & 170.000 & 48.000 & 820.000\\
5.300 & 48.000 & 0.140 & 21.000\\
3.600 & 820.000 & 21.000 & 0.004
\end{bmatrix} \times \num{e-3} \, .
\end{eqnarray}

Lastly, for the Manson-Succop model ($\bm{X} = [a_0, a_1, a_2, C]^{T}$), the outcomes were
\begin{eqnarray}
\bm{\mu}^{\rm MS} &=& [24.8, - 0.011, - 0.000019, 0.0289]^{T} \, ,\\
\bm{\Sigma}^{\rm MS} &=& \begin{bmatrix}
4.600 & 3.200 & 6.700 & - 3.600\\
3.200 & 26.000 & 160.000 & -35.000\\
6.700 & 160.000 & 6.900 & 1.700\\
-3.600 & -35.000 & 1.700 & 0.070
\end{bmatrix} \times \num{e-3} \, .
\end{eqnarray}

\subsection{Material rupture time: A statistical examination}

\begin{figure}
\centering
\includegraphics[width=0.45\textwidth]{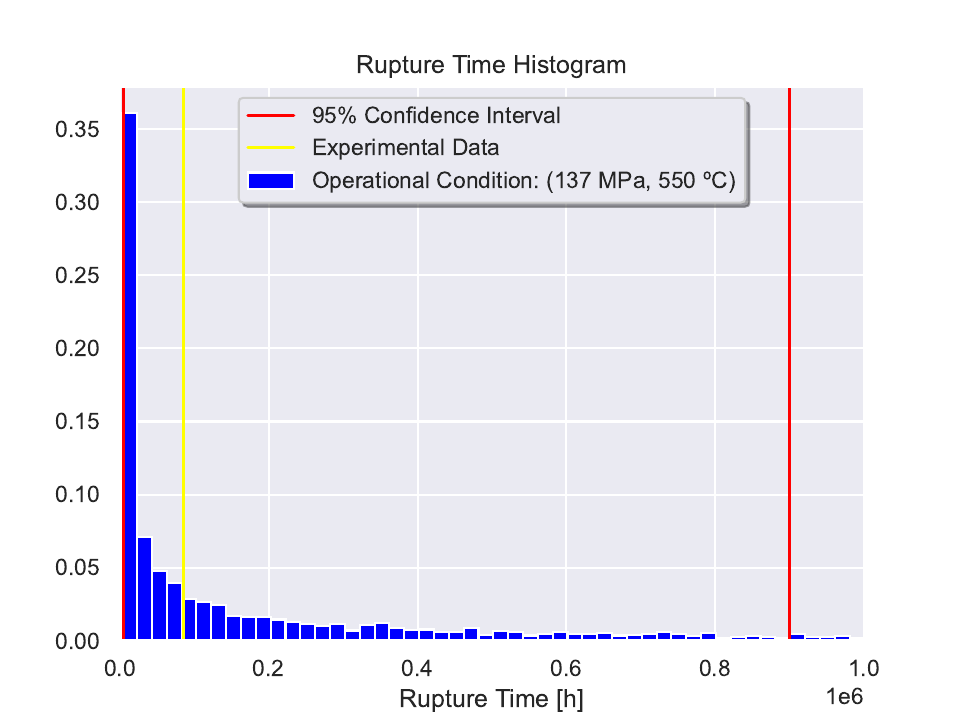}
\includegraphics[width=0.45\textwidth]{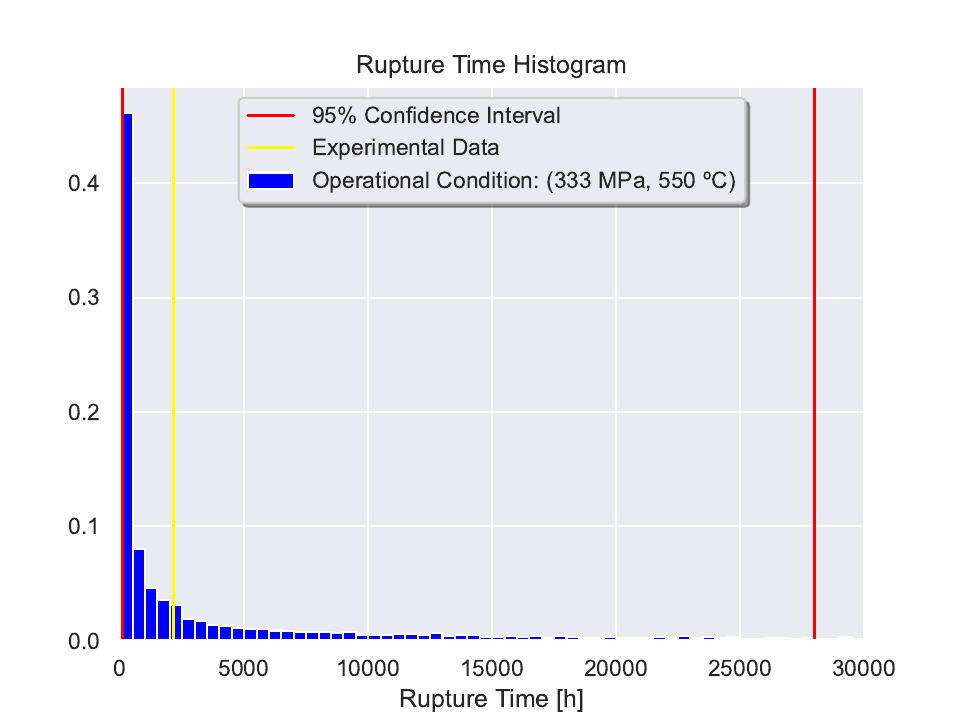}\\
\includegraphics[width=0.45\textwidth]{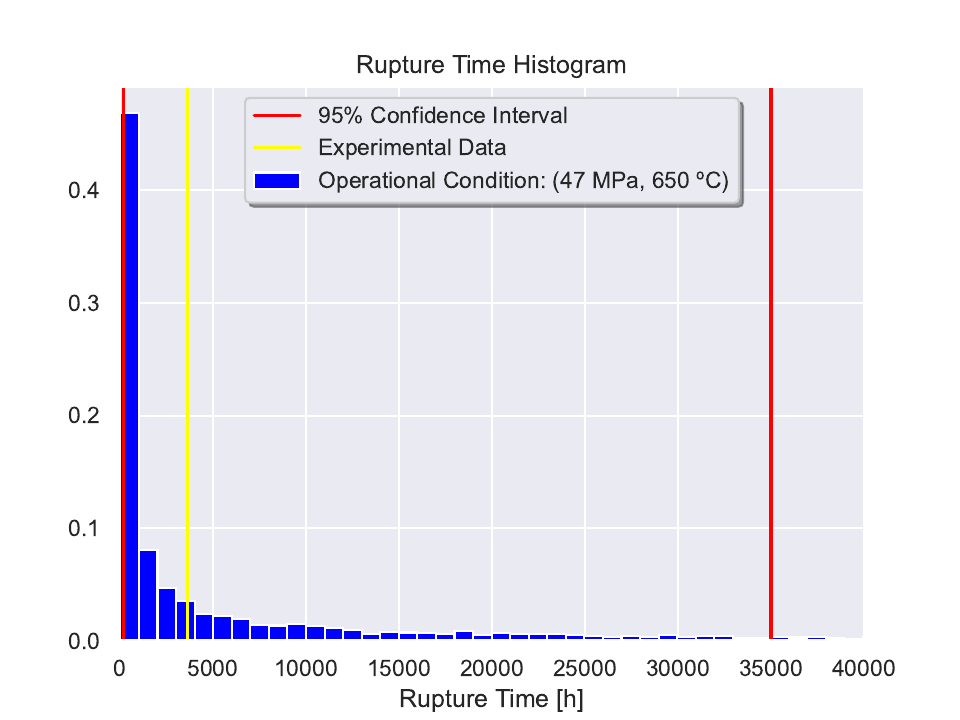}
\includegraphics[width=0.45\textwidth]{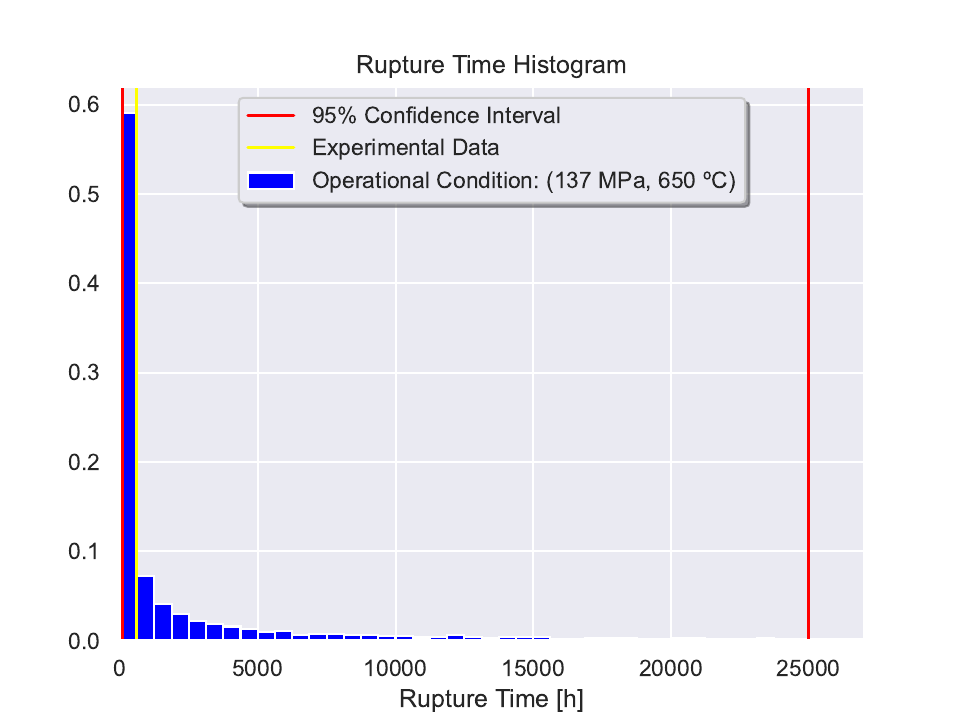}
\vspace{2mm}
\caption{Probability distribution of creep rupture time predicted by the Larson-Miller model under four operational conditions. Each histogram is based on 10,000 Monte Carlo simulations accounting for parametric uncertainty. Yellow vertical lines indicate experimental rupture time; red lines denote the 95\% confidence interval bounds. Note the pronounced skewness and long right tails—typical features in time-dependent degradation modeling}
\label{fig:HistPlm1}
\end{figure}

\begin{figure}
\centering
\includegraphics[width=0.45\textwidth]{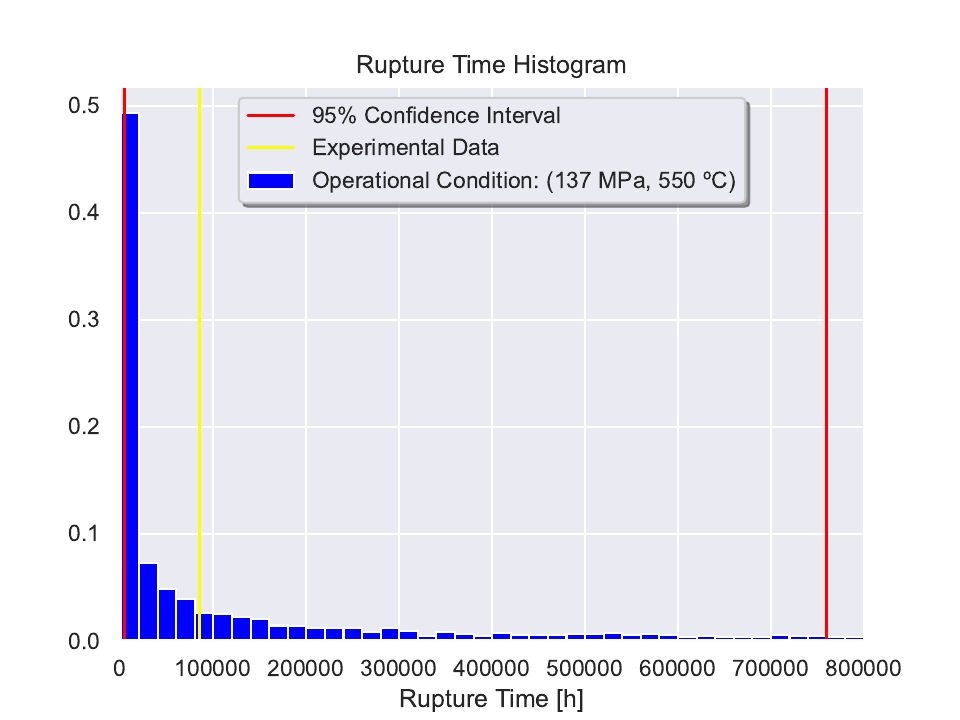}
\includegraphics[width=0.45\textwidth]{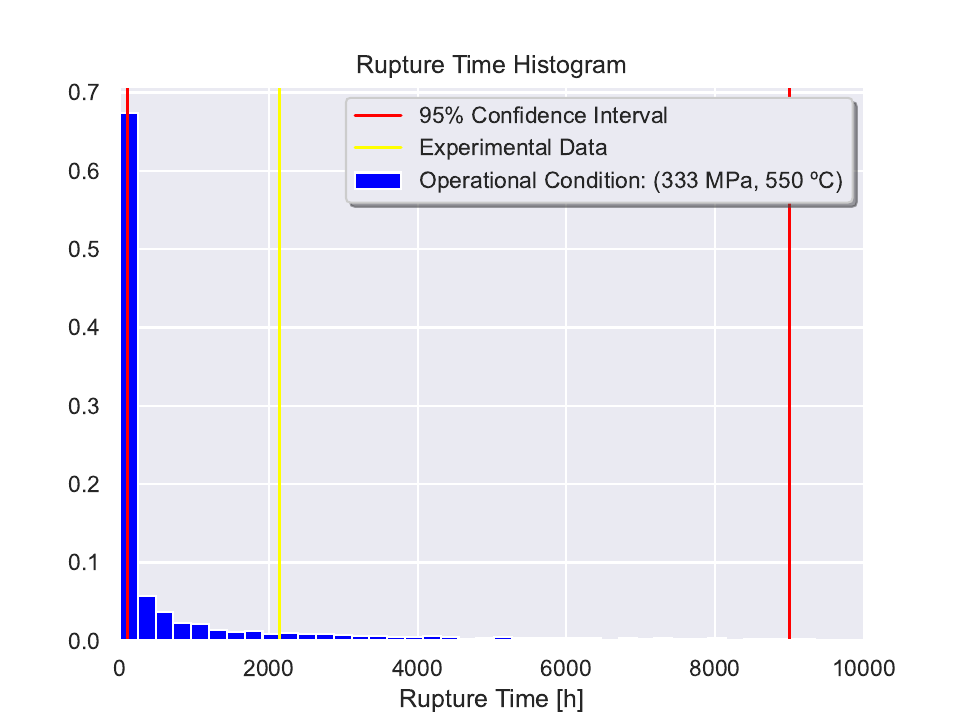}\\
\includegraphics[width=0.45\textwidth]{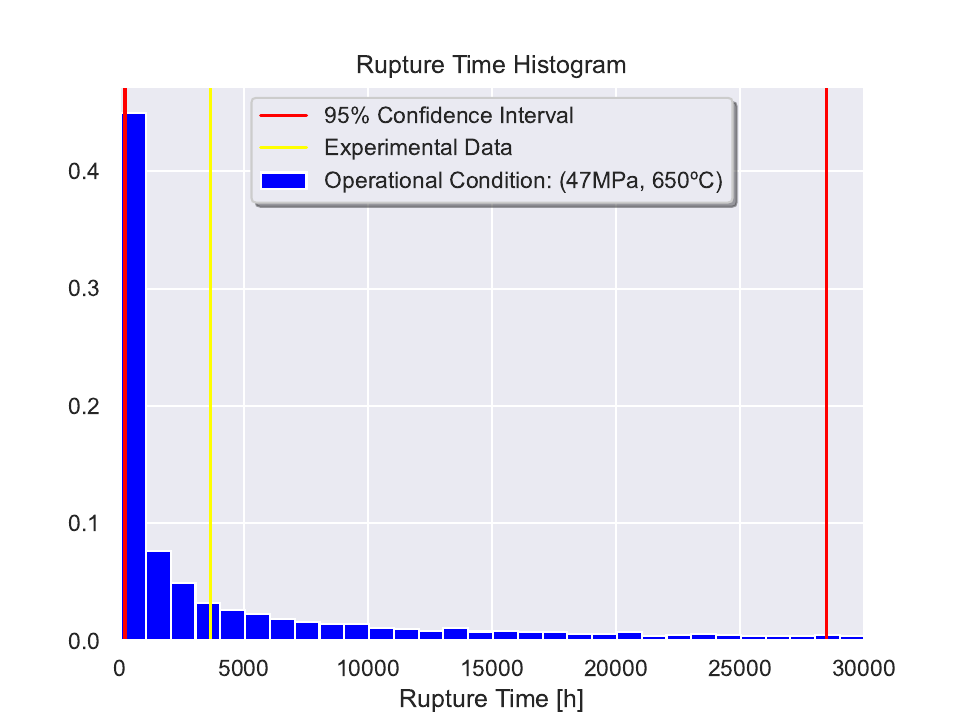}
\includegraphics[width=0.45\textwidth]{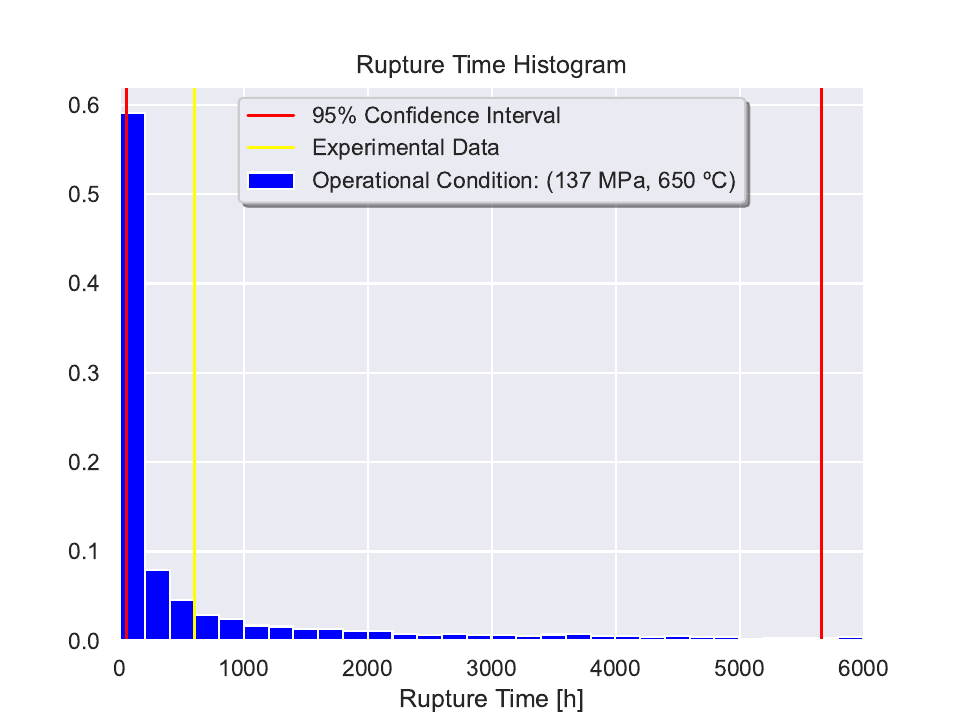}
\vspace{2mm}
\caption{Probability distribution of creep rupture time predicted by the Orr-Sherby-Dorn model under four operational conditions. Each histogram is based on 10,000 Monte Carlo simulations accounting for parametric uncertainty. Yellow vertical lines indicate experimental rupture time; red lines denote the 95\% confidence interval bounds. Note the pronounced skewness and long right tails—typical features in time-dependent degradation modeling.}
\label{fig:HistOSD1}
\end{figure}

\begin{figure}
\centering
\includegraphics[width=0.45\textwidth]{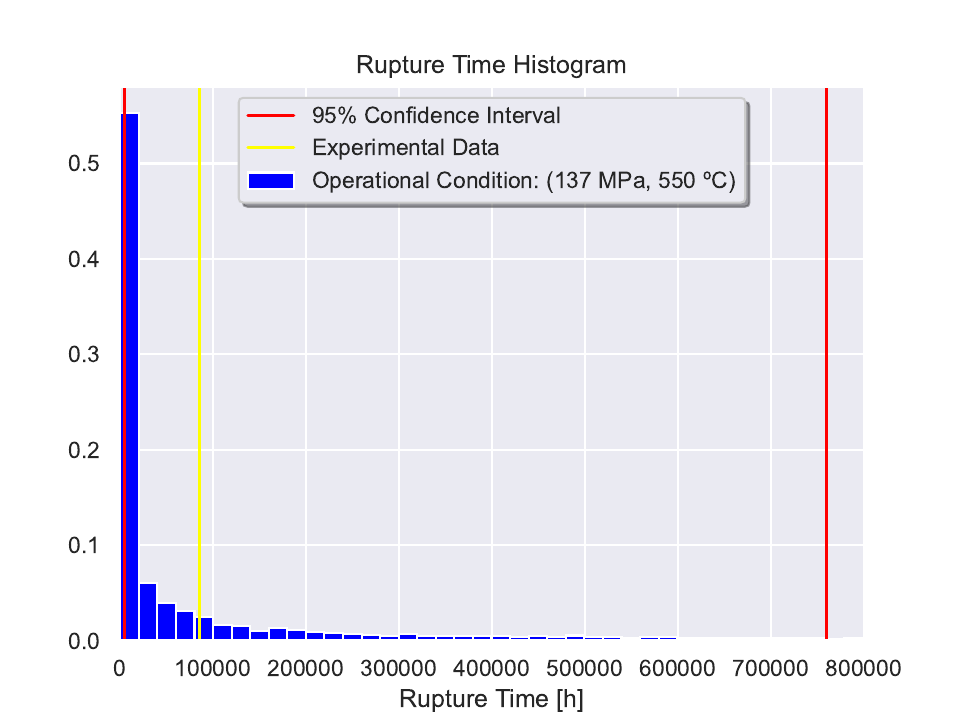}
\includegraphics[width=0.45\textwidth]{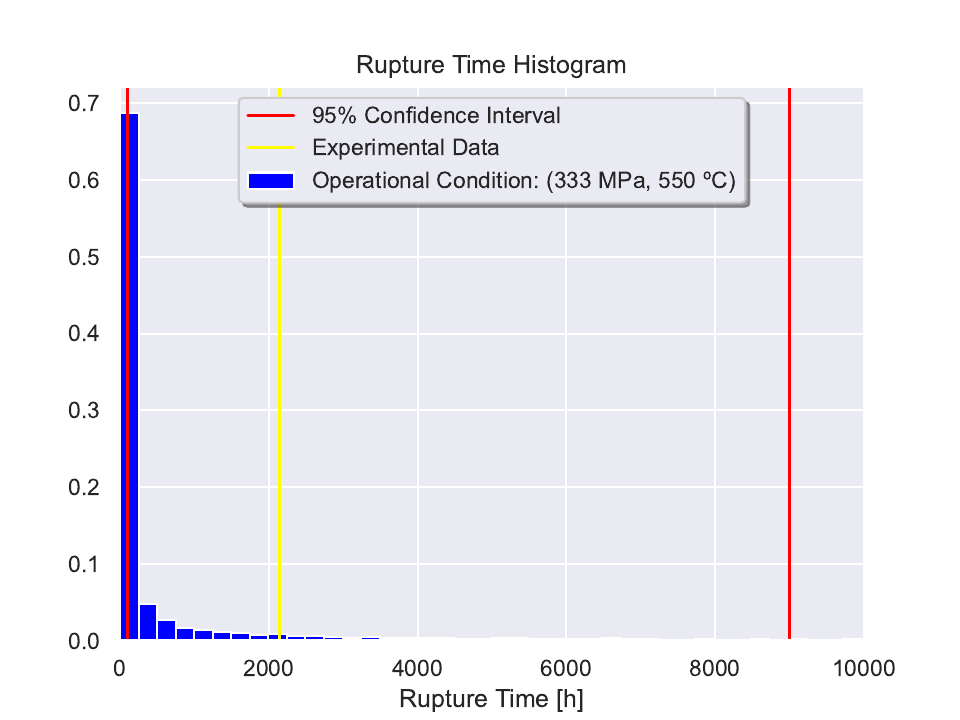}\\
\includegraphics[width=0.45\textwidth]{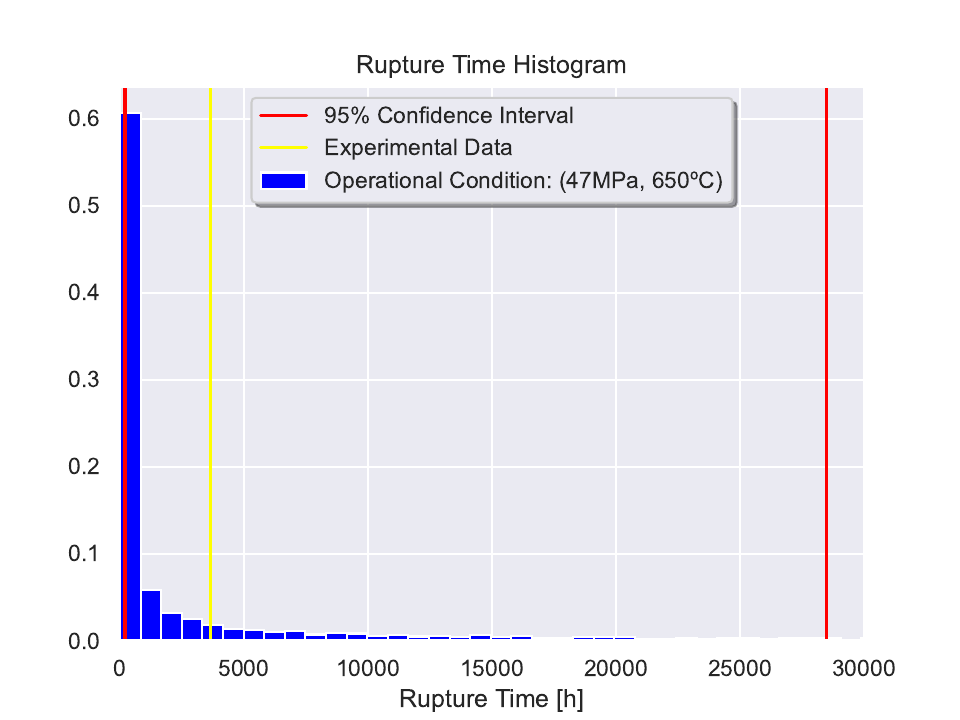}
\includegraphics[width=0.45\textwidth]{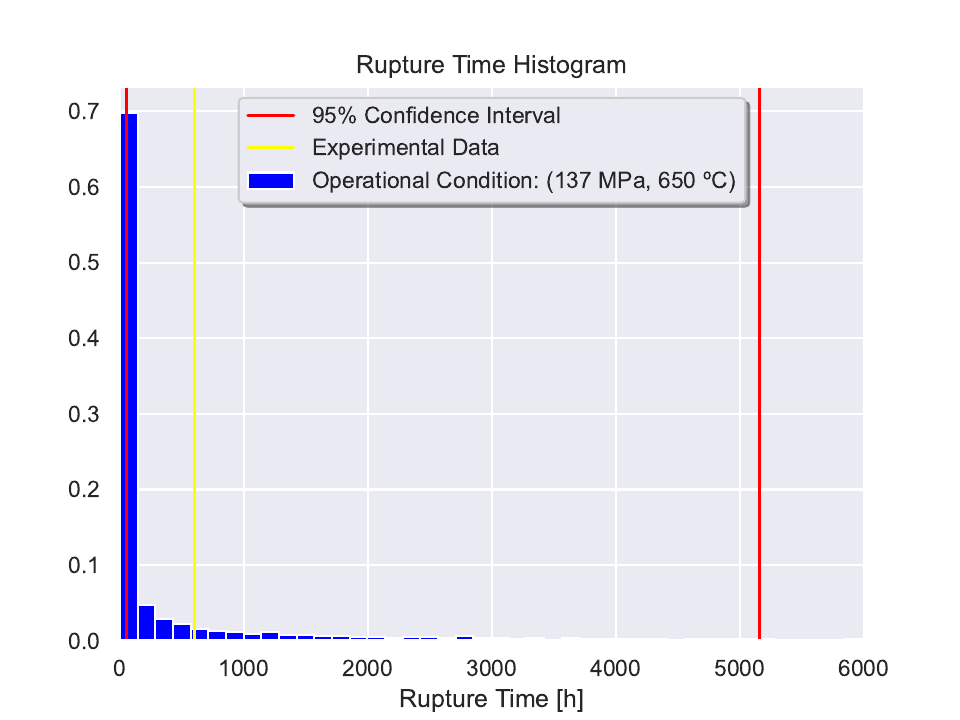}
\vspace{2mm}
\caption{Probability distribution of creep rupture time predicted by the Manson-Succop model under four operational conditions. Each histogram is based on 10,000 Monte Carlo simulations accounting for parametric uncertainty. Yellow vertical lines indicate experimental rupture time; red lines denote the 95\% confidence interval bounds. Note the pronounced skewness and long right tails—typical features in time-dependent degradation modeling.}
\label{fig:HistMS1}
\end{figure}

Figures \ref{fig:HistPlm1}, \ref{fig:HistOSD1}, and \ref{fig:HistMS1} show the empirical distributions of the creep rupture time $t_r$, obtained from 10,000 Monte Carlo samples for each parametric model under the four experimental conditions specified in Table \ref{table:EXP_conditions}. The experimental values (yellow lines) are consistently enclosed within the 95\% confidence intervals (red lines) predicted by the probabilistic models, validating their fidelity.

From an engineering reliability standpoint, these distributions mirror the statistical behavior commonly observed in fatigue life experiments, such as S–N or $\epsilon$-N data. The rupture time distributions exhibit strong right-skewness (skewness between 2.2 and 4.3) and pronounced kurtosis (ranging from 4.2 up to 19.8), as shown in Tables \ref{table:StatHistLM1} through \ref{table:StatHistMS1}. These high-order moments reflect the asymmetric dispersion and long-tailed behavior that often challenge fatigue engineers when dealing with lifetime variability in crack initiation or propagation.

The coefficients of variation (CoV) exceed 175\% in all cases—reaching over 300\% for the Orr-Sherby-Dorn model in Condition 2—confirming a substantial degree of uncertainty in the predicted lifetime. This is fully analogous to the scatter found in fatigue data, which stems from microstructural heterogeneities, surface finish, residual stress states, and test variability. Here, a similar stochasticity arises from thermal activation and material degradation mechanisms.

Despite this dispersion, the mean values from the probabilistic models are close to the experimental benchmarks, reinforcing the calibration effectiveness. However, it is the full distribution—rather than any single central estimate—that provides meaningful insight for structural reliability. In both creep and fatigue contexts, design for safety-critical components must account for rare but plausible outlier events, which reside in the distribution tails.

Moreover, the shape of the distributions appears to be relatively invariant with respect to the applied stress and temperature—a subtle but important observation. While the mean rupture time naturally decreases with more severe loading conditions, the persistence of heavy tails suggests that uncertainty in time-to-failure is intrinsic to the degradation process, not merely a function of operational parameters.

In summary, the probabilistic framework employed here not only quantifies the expected rupture time but also captures the broad spectrum of possible outcomes, offering a statistically grounded alternative to deterministic safety factors. The resulting insights and methodology are equally applicable to fatigue modeling, where similar probabilistic distributions govern life under cyclic loading, particularly in high-stakes applications such as aerospace, power generation, and nuclear systems.


\begin{table}
\centering
\caption{Experimental conditions applied in the study.}
\vspace{2mm}
\begin{tabular}{ccc}
\toprule
Condition & Stress {[}MPa{]} & Temperature {[}°C{]} \\
\midrule
1 & 137 & 550\\ 
2 & 333 & 550\\ 
3 & 47  & 650\\ 
4 & 137 & 650\\
\bottomrule
\end{tabular}
\label{table:EXP_conditions}
\end{table}


\begin{table}
\centering
\caption{Calculated statistics for creep rupture time under the Larson-Miller parametric model for four experimental conditions as detailed in Table~\ref{table:EXP_conditions}.}
\vspace{2mm}
\begin{tabular}{@{}lcccc@{}}
\toprule
 Condition & 1 & 2 &  3 & 4\\
\midrule
Mean               & 134,810h & 1,774h & 10,574h & 106h \\ 
Standard deviation & 337,025h & 3,140h & 21,888h & 240h \\ 
Skewness           & 3.0      & 2.2    & 2.2     & 2.9  \\ 
Kurtosis           & 9.2      & 4.2    & 4.2     & 8.7  \\ 
Coefficient of variation & 225\%    & 177\%   & 207\% & 219\%    \\ \bottomrule
\end{tabular}
\label{table:StatHistLM1}
\end{table}

\begin{table}
\centering
\caption{Calculated statistics for creep rupture time under the Orr-Sherby-Dorn parametric model for four experimental conditions as detailed in Table~\ref{table:EXP_conditions}.}
\vspace{2mm}
\begin{tabular}{@{}lcccc@{}}
\toprule
Condition & 1 & 2 &  3 & 4\\
\midrule
Mean               & 103,986h & 520h & 9,915h & 704h \\ 
Standard deviation & 200,718h & 1,585h & 19,677h & 1,650h \\ 
Skewness           & 2.4      & 4.3    & 2.6     & 3.2  \\ 
Kurtosis           & 5.6      & 19.8    & 6.2     & 10.4  \\ 
Coefficient of variation & 190\%    & 304\%   & 198\% & 234\%    \\ \bottomrule
\end{tabular}
\label{table:StatHistOSD1}
\end{table}

\begin{table}
\centering
\caption{Calculated statistics for creep rupture time under the Manson-Succop parametric model for four experimental conditions as detailed in Table~\ref{table:EXP_conditions}.}
\vspace{2mm}
\begin{tabular}{@{}lcccc@{}}
\toprule
Condition & 1 & 2 &  3 & 4\\
\midrule
Mean               & 202,235h & 2,694h & 8,098h & 691h \\ 
Standard deviation & 410,148h & 7,583h & 18,079h & 1,726h \\
Skewness           & 2.6      & 3.7    & 2.8     & 3.2  \\ 
Kurtosis           & 6.0      & 14.5    & 7.9     & 10.2  \\ 
Coefficient of variation & 202\%    & 280\%   & 223\% & 249\%    \\ \bottomrule
\end{tabular}
\label{table:StatHistMS1}
\end{table}



\subsection{Identifying the optimal creep model}

The appropriateness of the probabilistic model, where the mean value of creep rupture time closely mirrors the respective experimental value, varies depending on the specific experimental condition. For example, the Orr-Sherby-Dorn parametric model yielded the minimum relative error (1.5\%) for experimental condition 2. Conversely, the Larson-Miller parametric model provided the smallest relative error (11\%) for experimental condition 4. In general, the Orr-Sherby-Dorn parametric model proved to be more efficient at lower temperatures, as indicated by experimental conditions 1 and 2, despite displaying the most considerable dispersion (highest coefficient of variation) for experimental condition 2. Conversely, the Larson-Miller parametric model performed least effectively for operational conditions 3 and 4 (at higher temperatures), with experimental values of creep rupture time for these two conditions aligning closer to the lower limit of the 95\% confidence interval calculated with the probabilistic Larson-Miller creep model. Hence, the probabilistic creep models exhibited superior performance under lower temperature conditions, a conclusion supported by a comparison of the relative errors between mean values and measured data.

The model selection outcomes are collated in Table \ref{table:IFC}, where the Akaike Information Criterion (AIC) and the Bayesian Information Criterion (BIC) are computed for the three probabilistic creep models studied in this research. It's pertinent to note that the AIC and BIC values shown in Table \ref{table:IFC} were computed by evaluating the likelihood function at the mean value $\bm{\mu}$.

\begin{table}[h]
\centering
\caption{Calculated Akaike and Bayesian Information Criteria for the three probabilistic creep models.}
\begin{tabular}{lcc}
\toprule
Probabilistic Model & AIC & BIC \\
\midrule
Larson-Miller & 298.25 & 298.35\\ 
Orr-Sherby-Dorn & 367.46 & 370.53\\ 
Manson-Succop & 384.87  & 391.34\\
\bottomrule
\end{tabular}
\label{table:IFC}
\end{table}

AIC scores were marginally lower for the three parametric creep models, which was expected as the Akaike criterion is less stringent in penalizing model complexity compared to the Bayesian criterion. The probabilistic Larson-Miller parametric model obtained the lowest values of both AIC and BIC. Its governing parameter (the Larson-Miller parameter) and mechanical stress maintain the simplest relationship (via an affine function). Conversely, the Orr-Sherby-Dorn and Manson-Succop parameters are quadratic functions of mechanical stress, adding complexity due to the increase in model parameters. Although the probabilistic Larson-Miller model is the least complex and exhibits superior predictive performance, it does not outperform the other two models significantly, as evidenced by the comparable AIC and BIC values obtained across all three probabilistic creep models.

\section{Conclusion}

This study introduces a probabilistic framework applied to the structural integrity of high-temperature metals subjected to the phenomenon of creep. It utilizes three well-established parametric models to predict creep rupture time, thereby demonstrating the utility of the framework.

An additional crucial contribution of this work is the enhancement of connections between probabilistic and statistical methods within the structural integrity community. This calls for a committed effort towards mastering probabilistic and statistical concepts and establishing a uniform understanding of these techniques within the community. The authors are optimistic that the broader dissemination of these methodologies will encourage their widespread adoption. This increased acceptance will foster more contributors, facilitating the maturity of this fresh perspective, thereby paving the way for the development of unified probabilistic frameworks to confront any structural integrity challenges.

Probabilistic methods should not be viewed as replacements for traditional deterministic methods. They should be considered a more comprehensive approach, introducing a fresh perspective on tackling structural integrity issues. Probabilistic methods can incorporate all model uncertainties without resorting to excessive conservatism due to gaps in knowledge. Hence, they can be seen as an extension beyond the conventional deterministic methods, which are commonly disseminated through statistical techniques and physical models for failure prevention, all the while being bolstered by advancements in scientific computation. A crucial aspect of applying a probabilistic framework to any structural integrity problem involves a deep understanding of the physics underlying the phenomena leading to structural deterioration, along with the information provided by observed data. Together, these two sources of information facilitate the creation of physically meaningful statistical models for both input parameters and output quantities associated with the structure's reliability and integrity.

\section*{Acknowledgements}

The authors express their gratitude for the financial support received from the Electric Power Research Center (CEPEL), Coordination for the Improvement of Higher Education Personnel - Brazil (CAPES) - Finance Code 001, National Council for Scientific and Technological Development (CNPq), grant 305476/2022-0, and the Carlos Chagas Filho Research Foundation of Rio de Janeiro State (FAPERJ), grants: 211.037/2019, and 204.477/2024.

\section*{CRediT authorship contribution statement}

\textbf{VM:} Conceptualization, Methodology, Software implementation, Data post-processing, Formal analysis, Investigation, Writing – original draft, Writing – review \& editing. \textbf{CFTM:} Conceptualization, Methodology, Formal analysis, Fund acquisition, Supervision, Writing – review \& editing. \textbf{AC:} Conceptualization, Methodology, Formal analysis, Fund acquisition, Supervision, Writing – review \& editing.

\section*{Declaration of competing interest}

The authors declare that they have no known competing financial interests or personal relationships that could have appeared to influence the work reported in this paper.

\section*{Disclaimer}

This manuscript has undergone comprehensive grammatical review and enhancement using artificial intelligence-powered tools, including Grammarly and ChatGPT. However, the authors maintain full responsibility for the original language and wording.


\begin{thebibliography}{10}
\expandafter\ifx\csname url\endcsname\relax
  \def\url#1{\texttt{#1}}\fi
\expandafter\ifx\csname urlprefix\endcsname\relax\def\urlprefix{URL }\fi
\expandafter\ifx\csname href\endcsname\relax
  \def\href#1#2{#2} \def\path#1{#1}\fi

\bibitem{Dowling2012}
N.~E. Dowling, Mechanical Behavior of Materials, 4th Edition, Prentice Hall,
  2012.

\bibitem{Dias2023p185}
F.~Dias, L.~F. {Paullo Muñoz}, D.~Roehl, A numerical model for basic creep of
  concrete with aging and damage on beams, Applied Mathematical Modelling 121
  (2023) 185--203.
\newblock \href {https://doi.org/10.1016/j.apm.2023.04.018}
  {\path{doi:10.1016/j.apm.2023.04.018}}.

\bibitem{Evans2011p2838}
M.~Evans, Obtaining confidence limits for safe creep life in the presence of
  multi batch hierarchical databases: {A}n application to {18Cr–12Ni–Mo}
  steel, Applied Mathematical Modelling 35~(6) (2011) 2838--2854.
\newblock \href {https://doi.org/10.1016/j.apm.2010.11.072}
  {\path{doi:10.1016/j.apm.2010.11.072}}.

\bibitem{Gao2021p435}
Y.~Gao, D.~Yin, A full-stage creep model for rocks based on the variable-order
  fractional calculus, Applied Mathematical Modelling 95 (2021) 435--446.
\newblock \href {https://doi.org/10.1016/j.apm.2021.02.020}
  {\path{doi:10.1016/j.apm.2021.02.020}}.

\bibitem{Kamdem2023p624}
T.~C. Kamdem, K.~G. Richard, T.~B\'{e}da, New description of the mechanical
  creep response of rocks by fractional derivative theory, Applied Mathematical
  Modelling 116 (2023) 624--635.
\newblock \href {https://doi.org/10.1016/j.apm.2022.11.036}
  {\path{doi:10.1016/j.apm.2022.11.036}}.

\bibitem{Wang2020p37}
L.~ye~Wang, F.~xi~Zhou, Analysis of elastic-viscoplastic creep model based on
  variable-order differential operator, Applied Mathematical Modelling 81
  (2020) 37--49.
\newblock \href {https://doi.org/10.1016/j.apm.2019.12.007}
  {\path{doi:10.1016/j.apm.2019.12.007}}.

\bibitem{Roya2010}
\protect{N. Roya, S. Boseb and R. Ghoshc}, Stochastic aspects of evolution of
  creep damage in austenitic stainless steel, Materials Science and Engineering
  A 527 (2010) 4810--4817.
\newblock \href {https://doi.org/10.1016/j.msea.2010.04.013}
  {\path{doi:10.1016/j.msea.2010.04.013}}.

\bibitem{Smith:2013}
R.~C. Smith, Uncertainty Quantification: Theory, Implementation and
  Applications, SIAM, 2013.

\bibitem{soize2017}
C.~Soize, Uncertainty Quantification: An Accelerated Course with Advanced
  Applications in Computational Engineering, Springer, 2017.

\bibitem{Phanetal2017}
\protect{V.\ -T.\ Phan, X.\ Zhang, Y.\ Li and C.\ Oskay}, \protect{Microscale
  modeling of creep deformation and rupture in Nickel-based superalloy IN 617
  at high temperature}, Mechanics of Materials 114 (2017) 215--227.
\newblock \href {https://doi.org/10.1016/j.mechmat.2017.08.008}
  {\path{doi:10.1016/j.mechmat.2017.08.008}}.

\bibitem{Praveenetal2019}
\protect{C.\ Praveen, J.\ Christopher, V.\ Ganesan, G.\ V.\ Prasad Reddy, G.\
  Sasikala and S.\ K.\ Albert}, \protect{Constitutive modelling of transient
  and steady state creep behaviour of type 316LN austenitic stainless steel},
  Mechanics of Materials 137 (2019) 103--122.
\newblock \href {https://doi.org/10.1016/0045-7949(94)E0253-X}
  {\path{doi:10.1016/0045-7949(94)E0253-X}}.

\bibitem{Zentuti2017}
\protect{N. A. Zentuti, J. D. Booker and R. A. W. Bradford}, A review of
  probabilistic techniques: towards developing a probabilistic lifetime
  methodology in the creep regime, Materials at High Temperatures 34 (2017)
  333--341.
\newblock \href {https://doi.org/10.1080/09603409.2017.1371933}
  {\path{doi:10.1080/09603409.2017.1371933}}.

\bibitem{Gu2022p106677}
H.-H. Gu, R.-Z. Wang, S.-P. Zhu, X.-W. Wang, D.-M. Wang, G.-D. Zhang, Z.-C.
  Fan, X.-C. Zhang, S.-T. Tu, Machine learning assisted probabilistic
  creep-fatigue damage assessment, International Journal of Fatigue 156 (2022)
  106677.
\newblock \href {https://doi.org/10.1016/j.ijfatigue.2021.106677}
  {\path{doi:10.1016/j.ijfatigue.2021.106677}}.

\bibitem{Nie2025p108732}
W.-R. Nie, H.-H. Gu, X.-C. Zhang, S.-T. Tu, R.-Z. Wang, Hybrid-driven
  probabilistic damage assessment of creep-fatigue-oxidation interaction,
  International Journal of Fatigue 192 (2025) 108732.
\newblock \href {https://doi.org/10.1016/j.ijfatigue.2024.108732}
  {\path{doi:10.1016/j.ijfatigue.2024.108732}}.

\bibitem{MohammadLouetal2020}
\protect{B.\ S.\ Mohammad Lou, M.\ Pourgol-Mohamma and M.\ Yazdani},
  Probabilistic life assessment of gas turbine blade alloys under creep,
  International Journal of Reliability, Risk Safety: Theory and Application 3
  (2020) 9--17.
\newblock \href {https://doi.org/10.30699/IJRRS.3.2.2}
  {\path{doi:10.30699/IJRRS.3.2.2}}.

\bibitem{zhaoetal2009}
\protect{J.\ Zhao, D.\ M.\ Li, J.\ S.\ Zhang, W.\ Feng and Y.\ Fang},
  \protect{Introduction of SCRI model for creep rupture life assessment},
  International Journal of Pressure Vessels and Piping 86 (2009) 599--603.
\newblock \href {https://doi.org/10.1016/j.ijpvp.2009.04.004}
  {\path{doi:10.1016/j.ijpvp.2009.04.004}}.

\bibitem{Dias2019}
\protect{J.\ P.\ Dias, S.\ Ekwaro-Osire, A.\ Cunha Jr, S.\ Dabetwar, A.\
  Nispel, F.\ M.\ Alemayehu and H.\ B.\ Endeshaw}, Parametric probabilistic
  approach for cumulative fatigue damage using double linear damage rule
  considering limited data, International Journal of Fatigue 127 (2019)
  246--258.
\newblock \href {https://doi.org/10.1016/j.ijfatigue.2019.06.011}
  {\path{doi:10.1016/j.ijfatigue.2019.06.011}}.

\bibitem{AI2019165}
Y.~Ai, S.~Zhu, D.~Liao, J.~Correia, C.~Souto, A.~{De Jesus}, B.~Keshtegar,
  Probabilistic modeling of fatigue life distribution and size effect of
  components with random defects, International Journal of Fatigue 126 (2019)
  165--173.
\newblock \href {https://doi.org/10.1016/j.ijfatigue.2019.05.005}
  {\path{doi:10.1016/j.ijfatigue.2019.05.005}}.

\bibitem{Liu2023p107734}
X.-W. Liu, D.-G. Lu, Uncertainties quantification of fatigue load mixture model
  using hierarchical bayesian models, International Journal of Fatigue 174
  (2023) 107734.
\newblock \href {https://doi.org/10.1016/j.ijfatigue.2023.107734}
  {\path{doi:10.1016/j.ijfatigue.2023.107734}}.

\bibitem{Meggiolaro2023p107315}
M.~A. Meggiolaro, J.~T.~P. de~Castro, Probabilistic stress and strain-life
  fatigue crack initiation models with mean stress effects and life-dependent
  scatter considering runouts, International Journal of Fatigue 167 (2023)
  107315.
\newblock \href {https://doi.org/10.1016/j.ijfatigue.2022.107315}
  {\path{doi:10.1016/j.ijfatigue.2022.107315}}.

\bibitem{Ball2024p108569}
D.~L. Ball, M.~A. Ryan, Uncertainty quantification in residual stress affected
  fatigue crack growth life, International Journal of Fatigue 189 (2024)
  108569.
\newblock \href {https://doi.org/10.1016/j.ijfatigue.2024.108569}
  {\path{doi:10.1016/j.ijfatigue.2024.108569}}.

\bibitem{Dabetwar2021p021004}
S.~Dabetwar, S.~Ekwaro-Osire, J.~P. Dias, Fatigue damage diagnostics of
  composites using data fusion and data augmentation with deep neural networks,
  Journal of Nondestructive Evaluation, Diagnostics and Prognostics of
  Engineering Systems 5~(2) (2021) 021004.
\newblock \href {https://doi.org/10.1115/1.4051947}
  {\path{doi:10.1115/1.4051947}}.

\bibitem{Deng2025p108647}
X.~Deng, S.-P. Zhu, L.~Wang, C.~Luo, S.~Fu, Q.~Wang, Probabilistic framework
  for strain-based fatigue life prediction and uncertainty quantification using
  interpretable machine learning, International Journal of Fatigue 190 (2025)
  108647.
\newblock \href {https://doi.org/10.1016/j.ijfatigue.2024.108647}
  {\path{doi:10.1016/j.ijfatigue.2024.108647}}.

\bibitem{Zhangetal2024}
\protect{Z.\ Zhang, X.\ Wang, Z.\ Li, X.\ Xia, Y.\ Chen, T.\ Zhang, H.\ Zhang,
  Z.\ Yang, X.\ Zhang and J.\ Gong}, Machine learning-assisted probabilistic
  creep life assessment for high-temperature superheater outlet header
  considering material uncertainty, International Journal of Pressure Vessels
  and Piping 209 (2024) 1--12.
\newblock \href {https://doi.org/10.1016/j.ijpvp.2024.105211}
  {\path{doi:10.1016/j.ijpvp.2024.105211}}.

\bibitem{HarlowDelph1995}
\protect{D.\ G.\ Harlow and T.\ J.\ Delph}, \protect{A computational
  probabilistic model for creep-damaging solids}, Computers \& Structures 54
  (1995) 161--166.
\newblock \href {https://doi.org/10.1016/j.ijpvp.2021.104446}
  {\path{doi:10.1016/j.ijpvp.2021.104446}}.

\bibitem{HossainStewart2021}
\protect{M.\ A.\ Hossain and C.\ M.\ Stewart}, \protect{A probabilistic creep
  model incorporating test condition, initial damage, and material property
  uncertainty}, International Journal of Pressure Vessels and Piping 193 (2021)
  104446.
\newblock \href {https://doi.org/10.1016/j.ijpvp.2021.104446}
  {\path{doi:10.1016/j.ijpvp.2021.104446}}.

\bibitem{Ribeiro2021p1184}
J.~G. {Telles Ribeiro}, J.~T.~P. de~Castro, M.~A. Meggiolaro, Modeling concrete
  and polymer creep using fractional calculus, Journal of Materials Research
  and Technology 12 (2021) 1184--1193.
\newblock \href {https://doi.org/10.1016/j.jmrt.2021.03.007}
  {\path{doi:10.1016/j.jmrt.2021.03.007}}.

\bibitem{TellesRibeiro2025p20240861}
J.~G. {Telles Ribeiro}, A.~{Cunha~Jr}, Advanced creep modelling for polymers:
  {A} variable-order fractional calculus approach, Proceedings of the Royal
  Society A 481 (2025) 20240861.
\newblock \href {https://doi.org/10.1098/rspa.2024.0861}
  {\path{doi:10.1098/rspa.2024.0861}}.

\bibitem{Miller1952}
\protect{F.\ Larson and J.\ Miller}, A time-temperature relationship for
  rupture and creep stresses, Transactions of the American Society of
  Mechanical Engineers 74 (1952) 765--775.

\bibitem{Choudharyetal2014}
\protect{B.\ K.\ Choudhary, W.\ -G.\ Kim, M.\ D.\ Mathew, J.\ Jang, T.\
  Jayakumar and Y.\ -H.\ Jeong}, On the reliability assessment of creep life
  for grade 91 steel, Procedia Engineering 86 (2014) 335--341.
\newblock \href {https://doi.org/10.1016/j.proeng.2014.11.046}
  {\path{doi:10.1016/j.proeng.2014.11.046}}.

\bibitem{Ayubalietal2021}
\protect{A.\ A.\ Ayubali, A.\ Singh, B.\ P.\ Shanmugavel and K.\ A.\
  Padmanabhan}, A phenomenological model for predicting long-term high
  temperature creep life of materials from short-term high temperature creep
  test data, International Journal of Mechanical Sciences 202-203 (2021)
  106505.
\newblock \href {https://doi.org/10.1016/j.ijmecsci.2021.106505}
  {\path{doi:10.1016/j.ijmecsci.2021.106505}}.

\bibitem{Monkman1956}
\protect{F.\ Monkman and N.\ Grant}, An empirical relationship between rupture
  life and minimum creep rate in creep rupture tests, ASTM Proceedings 56
  (1956) 593--620.

\bibitem{Gilbert2007}
J.~Gilbert, Z.~Long, S.~Ningileri, Application of Time-Temperature-Stress
  Parameters to High Temperature Performance of Aluminium Alloys, The Minerals,
  Metals \& Materials Society, 2007.

\bibitem{Orr1954}
\protect{R.\ Orr, O.\ Sherby and J.\ Dorn}, Correlation of rupture data for
  metals at elevatnology, Institue of Engineering Research, Univ. of Calif.,
  Berkeley (1954).
\newblock \href {https://doi.org/10.2172/4425999} {\path{doi:10.2172/4425999}}.

\bibitem{Carreker1950}
\protect{R.\ Carreker}, Plastic flow of platinum wires., Journal of Applied
  Physics 21 (1950) 1289–1296.
\newblock \href {https://doi.org/10.1063/1.1699593}
  {\path{doi:10.1063/1.1699593}}.

\bibitem{Mullendore1963}
\protect{A.\ Mullendore, J.\ Dhosi and N.\ Grant}, Study of parameter
  techniques for the extrapolation of creep rupture properties, in conference
  proceedings 1963, Proceedings of the Institution of Mechanical Engineers. ~
  (1963) ~.

\bibitem{Allen1960}
\protect{N.\ Allen}, \protect{The Extrapolation of Creep Tests, A Review of
  Recent Opinion}, Institute of Metals, London, UK (1960).

\bibitem{Manson1956}
\protect{S.\ Manson and G.\ Succop}, Stress-rupture properties of inconel 700
  and correlation on the basis of several time-temperature parameters, ASTM STP
  (1956) 40--46.

\bibitem{Zharkova2003}
\protect{N.\ Zharkova and L.\ Botvina}, Estimate of the life of a material
  under creep conditions in the phase transition theory, ASTM 391 (2003)
  334--336.

\bibitem{Abdallahetal2018}
\protect{Z.\ Abdallah, K.\ Perkins and C.\ Arnold}, Creep lifing models and
  techniques, in: \protect{T.\ A.\ Tański, M.\ Sroka and A.\ Zieliński}
  (Ed.), Creep, IntechOpen, Chapter 7, 2018, pp. 115--149.
\newblock \href {https://doi.org/10.5772/intechopen.71826}
  {\path{doi:10.5772/intechopen.71826}}.

\bibitem{Yang2025}
S.~Yang, D.~Meng, A.~D{\'\i}az, H.~Yang, X.~Su, A.~M.~d. Jesus, Probabilistic
  modeling of uncertainties in reliability analysis of mid- and high-strength
  steel pipelines under hydrogen-induced damage, International Journal of
  Structural Integrity 16~(1) (2025) 39--59.
\newblock \href {https://doi.org/10.1108/IJSI-10-2024-0177}
  {\path{doi:10.1108/IJSI-10-2024-0177}}.

\bibitem{Gong2024}
W.~Gong, X.-Y. Wang, X.~Wang, W.~Wang, Y.-L. Yang, Comparison of inelasticity
  creep failure evaluation codes for elevated-temperature non-nuclear pressure
  equipment, International Journal of Structural Integrity 15~(6) (2024)
  1153--1168.
\newblock \href {https://doi.org/10.1108/IJSI-07-2024-0101}
  {\path{doi:10.1108/IJSI-07-2024-0101}}.

\bibitem{Gharaibeh2024}
M.~A. Gharaibeh, J.~Wilde, Research on the creep response of lead-free die
  attachments in power electronics, International Journal of Structural
  Integrity 15~(4) (2024) 702--716.
\newblock \href {https://doi.org/10.1108/IJSI-01-2024-0005}
  {\path{doi:10.1108/IJSI-01-2024-0005}}.

\bibitem{Dias2018}
\protect{J.\ P.\ Dias, S.\ Ekwaro-Osire, A.\ Cunha Jr, F.\ M.\ Alemayehu, S.\
  Dabetwar and A.\ Nispel}, A parametric probabilistic approach to quantify
  uncertainties in a non-linear cumulative fatigue damage model considering
  limited data, in: Twelfth International Conference on Fatigue Damage of
  Structural Materials (ICFDSM 2018), Hyannis, United States, 2018.

\bibitem{Nispel2021}
\protect{A.\ Nispel, J.\ P.\ Dias, S.\ Ekwaro-Osire and A.\ Cunha Jr},
  Uncertainty quantification for fatigue life of offshore wind turbine
  structure, ASCE-ASME Journal of Risk and Uncertainty in Engineering Systems
  Part B: Mechanical Engineering 7 (2021) 040901.
\newblock \href {https://doi.org/10.1115/1.4051162}
  {\path{doi:10.1115/1.4051162}}.

\bibitem{Boyd2018}
\protect{S.\ Boyd and L.\ Vandenberghe}, Introduction to Applied Linear Algebra
  – Vectors, Matrices, and Least Squares, Cambridge University Press, 2018.

\bibitem{Brunton2016}
\protect{S.\ L.\ Brunton, J.\ L.\ Proctor and J.\ N.\ Kutz}, Discovering
  governing equations from data by sparse identification of nonlinear dynamical
  systems, Proceedings of the National Academy of Sciences 113 (2016)
  3932--3937.
\newblock \href {https://doi.org/10.1073/pnas.1517384113}
  {\path{doi:10.1073/pnas.1517384113}}.

\bibitem{brunton2022}
\protect{S.\ L.\ Brunton and J.\ N.\ Kutz}, Data-Driven Science and
  Engineering: Machine Learning, Dynamical Systems, and Control, 2nd Edition,
  Cambridge University Press, 2022.

\bibitem{Sudret2008}
\protect{B.\ Sudret}, Global sensitivity analysis using polynomial chaos
  expansions, Reliability Engineering and System Safety 93~(7) (2008) 964--979.
\newblock \href {https://doi.org/10.1016/j.ress.2007.04.002}
  {\path{doi:10.1016/j.ress.2007.04.002}}.

\bibitem{kroese2011}
\protect{D.\ P.\ Kroese, T.\ Taimre and Z.\ I.\ Botev}, Handbook of Monte Carlo
  Methods, Wiley, 2011.

\bibitem{cunhajr2014p1355}
A.~{Cunha~Jr}, R.~Nasser, R.~Sampaio, H.~Lopes, K.~Breitman, Uncertainty
  quantification through {M}onte {C}arlo method in a cloud computing setting,
  Computer Physics Communications 185 (2014) 1355--1363.
\newblock \href {https://doi.org/https://doi.org/10.1016/j.cpc.2014.01.006}
  {\path{doi:https://doi.org/10.1016/j.cpc.2014.01.006}}.

\bibitem{ghanem2003}
\protect{R.\ G.\ Ghanem and P.\ D.\ Spanos}, Stochastic Finite Elements: A
  Spectral Approach, 2nd Edition, Dover Publications, 2003.

\bibitem{xiu2010}
D.~Xiu, Numerical Methods for Stochastic Computations: A Spectral Method
  Approach, Princeton University Press, 2010.

\bibitem{UQLab}
\protect{S.\ Marelli and B.\ Sudret}, {UQLab}: a framework for uncertainty
  quantification in {MATLAB}, in: Proc. 2nd Int. Conf. on Vulnerability, Risk
  Analysis and Management {(ICVRAM2014)}, Liverpool, United Kingdom, 2014, p.
  2554–2563.
\newblock \href {https://doi.org/10.1061/9780784413609.257}
  {\path{doi:10.1061/9780784413609.257}}.

\bibitem{cunhajr_ekwaro2016}
A.~{Cunha~Jr}, {M}odeling and {Q}uantification of {P}hysical {S}ystems
  {U}ncertainties in a {P}robabilistic {F}ramework, in: S.~Ekwaro-Osire, A.~C.
  Gon\c{c}alves, F.~M. Alemayehu (Eds.), Probabilistic Prognostics and Health
  Management of Energy Systems, Springer, Cham, 2017, pp. 127--156.
\newblock \href {https://doi.org/10.1007/978-3-319-55852-3_8}
  {\path{doi:10.1007/978-3-319-55852-3_8}}.

\bibitem{wasserman2004}
L.~Wasserman, All of Statistics: A Concise Course in Statistical Inference,
  Springer, 2004.

\bibitem{Hastie2016}
\protect{T.\ Hastie, R.\ Tibshirani and J.\ Friedman}, The Elements of
  Statistical Learning, Springer, 2016.

\bibitem{Murphy2012}
\protect{T.\ Hastie, R.\ Tibshirani and J.\ Friedman}, Machine learning: A
  probabilistic perspective, The MIT Press, 2012.

\end{thebibliography}


\end{document}